\documentclass[11pt,a4paper]{article}          
\usepackage{jheppub}

\usepackage{mathtools}
\usepackage{mathrsfs}
\usepackage{physics}
\usepackage{braket}
\usepackage[noabbrev]{cleveref}
\usepackage{empheq}
\usepackage{blkarray}

\usepackage{tikz}
\usetikzlibrary{positioning,arrows}
\tikzset{>=stealth}
\usetikzlibrary{decorations.pathmorphing}

\title{Higher-form (quasi)hydrodynamics from holography: deformations and dualities}

\author{Andr\'{e} Oliveira Pinheiro} 

\affiliation{Department of Mathematics and Maxwell Institute for Mathematical Sciences, Heriot-Watt University, Edinburgh EH14 4AS, U.K.}

\emailAdd{ao2012@hw.ac.uk}

\abstract{We study the low-energy dynamics of systems with exact and approximate higher-form symmetries using Gauge-gravity duality. These symmetries are realised holographically via generalised Maxwell/Proca theories for massless/massive $p$-forms in AlAdS spacetimes. Double-trace deformations of the boundary theory are considered via appropriate boundary conditions. We compute thermal correlation functions in isotropic black brane backgrounds to characterise the near equilibrium regimes of the dual boundary theories. In the vanishing-mass limit, the theory exhibits a hydrodynamic regime for weak double-trace deformations (relative to a scale set by the temperature) and a quasihydrodynamic regime for strong deformations. Turning on the bulk mass gives rise instead to a triad of quasihydrodynamic regimes controlled by both the mass and the double-trace coupling. In general, we find the low-energy spectra to be constrained by pole collisions, emergent symmetries and duality relations, the latter originating in part from Hodge-type dualities in the bulk. For nonzero mass, there is an additional strong/weak duality of the double-trace couplings. We further show, in the low-density limit of background charge, that relevant deformations are necessary for stable diffusion of sufficiently high-dimensional charged objects.}

\keywords{Holography and Hydrodynamics, Global Symmetries, Gauge-Gravity Correspondence}

\arxivnumber{2509.26583}

\begin{document}
\maketitle
\flushbottom

\section{Introduction} \label{intro}

\paragraph{}Driven by the framework of \textit{generalised symmetries} \cite{Gaiotto:2014kfa}, the past decade has witnessed significant progress in our knowledge of (global) symmetries in physics \cite{Cordova:2022ruw}. Notably, such progress has not come from new theories with exotic symmetries, but rather from a deeper understanding of familiar theories and the structures they exhibit. In particular, generalised symmetries have proven instrumental in extending the Landau paradigm \cite{Landau:1937obd} to include deconfined phases of gauge theories, topologically ordered phases, etc. \cite{McGreevy:2022oyu}. Generalised symmetries are often classified under various labels, such as higher-form, higher-group or non-invertible symmetries. For a broad overview see \cite{Brennan:2023mmt,Bhardwaj:2023kri} and also \cite{Gomes:2023ahz,Luo:2023ive,Iqbal:2024pee} for discussions with an emphasis on applications.\footnote{\cite{Bhardwaj:2023kri} contains an extensive account of the precursors to \cite{Gaiotto:2014kfa}.} \\ \hspace*{1.75ex}
The advent of generalised symmetries led to their use in formulating bottom-up holographic theories \cite{Maldacena:1997re,Witten:1998qj,Gubser:1998bc}, namely in the context of magnetohydrodynamics \cite{Grozdanov:2017kyl,Hofman:2017vwr}. Their application shortly after to holographic descriptions of viscoelastic crystals \cite{Grozdanov:2018ewh,Armas:2019sbe} is also noteworthy. 
Parallel to this, there was a purely hydrodynamic study of systems with higher-form symmetries \cite{Grozdanov:2016tdf,Armas:2018ibg,Armas:2018atq,Glorioso:2018kcp,Armas:2018zbe,Delacretaz:2019brr} (in $d$ spacetime dimensions):
\begin{itemize}
	\item Crystals without topological defects \cite{Armas:2019sbe} --- given $\mathrm{n}$ equal to (less than) $d{-}1$, the elastic (smectic) phase of these crystals is characterised by the n'th product of magnetic $(d{-}2)$-form symmetries, denoted hereafter as magnetic$_{(d-2)}$;\footnote{In this work, ``$p$-form symmetry" refers to a copy with U$(1)$ symmetry group. Hence, ``products of symmetries" are associated with U$(1) \times \ldots \times$U$(1)$ groups.}
	\item Superfluids \cite{Delacretaz:2019brr} --- possessing electric$_{(0)} \times$magnetic$_{(d-2)}$ symmetries with a mixed t'Hooft anomaly;
	\item Polarised plasmas in $d=4$ --- this phase of electromagnetism at finite temperature \cite{Armas:2018zbe} is characterised by electric$_{(1)} \times$magnetic$_{(1)}$ symmetries. (Due to Debye screening, the magnetohydrodynamic phase is described solely by the magnetic$_{(1)}$ symmetry).
\end{itemize}
We distinguish between \textit{electric} and \textit{magnetic} $p$-form symmetries \cite{Cordova:2022ruw} based on the latter being associated with $(d{-}p{-}2)$-form Goldstones arising from spontaneous breaking of continuous symmetries. In crystals, it is translation invariance in spatial directions that is spontaneously broken and in superfluids/polarised plasmas it is the electric symmetry.

\begin{figure}[t]
	\centering
	\includegraphics[width=8.5cm,height=3.0cm]{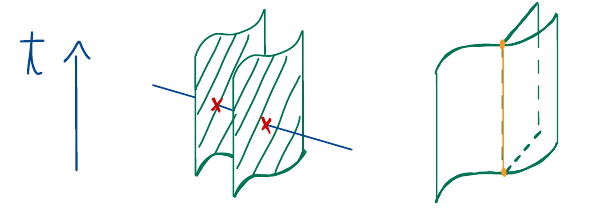}
	\caption{1-form symmetry with defects: on the left, time is indicated as running vertically; in the middle, two infinitely extended strings and their worldsheets are shown --- the 1-form symmetry is reflected in the fact that the number of intersections between the worldsheets and a codimension-2 hypersurface is topological; on the right, the symmetry is broken by a defect associated with a junction from which two strings emanate (or into which they merge).}
	\label{figdefects}
\end{figure}

\paragraph{}All the higher-form symmetries discussed above (and throughout the remainder of the paper) are continuous and are therefore tied to the conservation of higher-dimensional extended objects. A continuous $(p{-}1)$-form symmetry is associated with a conserved $p$-form current $\mathcal{O}$. When this symmetry is weakly broken, $\mathcal{O}$ is only approximately conserved such that
\begin{equation} \begin{aligned} \label{nonconseq}
	\mathrm{d} \ast \mathcal{O} = \ell \ast \tilde{\mathcal{O}} \, ,
\end{aligned} \end{equation}
where $\ell \ll 1$ and we call $\tilde{\mathcal{O}} \in \Omega^{p-1}$ the \textit{defect current} \cite{Armas:2023tyx}.\footnote{Conventions regarding exterior calculus are presented in appendix \ref{conventions_append}.} Note that, when $\ell$ vanishes, the conservation equation is recovered and the symmetry is said to be exact. In this case, the $p$-dimensional worldvolumes along which the extended charges propagate form a set of manifolds with no boundary. When $\ell \neq 0$, on the other hand, the worldvolumes form a set of hypersurfaces that are not necessarily manifolds nor boundaryless. The failure to meet these criteria is associated with junctions (as the higher-dimensional analogue of a branching structure) and boundaries, respectively, where $\tilde{\mathcal{O}} (x) \neq 0$. Junctions and boundaries are then the $(p{-}1)$-dimensional locus where local continuity is violated. When they lie within a codimension-1 timelike hypersurface, we call them \textit{defects} since, in this case, the junction/boundary corresponds to the worldvolume along which a $(p{-}2)$-dimensional imperfection of the extended charges propagates.\footnote{The \textit{imperfections} are themselves junctions or boundaries of the locus occupied by the extended charges. In appendix \ref{higherformsym}, we use them as motivation to \cref{nonconseq}.} The exterior derivative of \cref{nonconseq} implies a continuous $(p{-}2)$-form symmetry according to which the defect current is conserved. The defects are the worldvolumes of the objects charged under this symmetry and, therefore, they form a set of manifolds with no boundary. These ideas are illustrated in \cref{figdefects} (for $p=2$).

\paragraph{}The explicit breaking of higher-form symmetries arises from lower-dimensional imperfections (and the corresponding defects) in a variety of physical systems, including those discussed previously.
Dislocations in a crystal tend to form as its temperature is increased \cite{Friedel1964}. If their location is sparse enough, the magnetic$_{(d-2)}$ symmetries are weakly broken. Analogously to dislocations, vortices render the magnetic$_{(d-2)}$ symmetry of a superfluid approximate \cite{Putterman1974,Kleinert:1989kx}. Lastly, a polarised plasma is similar to free electromagnetism in the vacuum, in the sense that the electric$_{(1)}$ (magnetic$_{(1)}$) symmetry is explicitly broken when free electric charges (magnetic monopoles) are present.
Note that when a magnetic symmetry is broken, the Goldstone field becomes singular in a way that the associated physical observable (superfluid velocity, field strength, etc.) is still smooth --- we call this a \textit{multivalued Goldstone}.\footnote{These should not be mistaken for \textit{pseudo-Goldstone fields}, which occur when an approximate symmetry is spontaneously broken. In this case, the Goldstones acquire a small mass. For a general account of pseudo-Goldstones alongside many applications, see \cite{Delacretaz:2021qqu}. They have been studied holographically in \cite{Ammon:2021pyz} (which includes a study of a massive 1-form gauge field in the bulk).}

\paragraph{}Let us take a brief look at the (classical) hydrodynamics of higher-form symmetries. It is useful to assume, for simplicity, that we are in Minkowski flat space and consider separately the temporal and spatial components of \cref{nonconseq}:\footnote{Compared with \eqref{nonconseq}, we have rescaled $\ell$ according to $\ell \to (-1)^{d-p}  (d-p)! \ell$ in these equations.}
\begin{subequations} \begin{align} 
	\label{temporal}	\partial_{i_2} \mathcal{O}^{t i_2 ... i_p} + \ell \tilde{\mathcal{O}}^{t i_3 ... i_p} & = 0 \\
	\label{0.10} \partial_{t} \mathcal{O}^{t i_2 ... i_p} + \partial_{i_1} \mathcal{O}^{i_1 ... i_p} & = \ell \tilde{\mathcal{O}}^{i_2 ... i_p} \, .
\end{align} \end{subequations}
The temporal components (on top) are absent for an ordinary $0$-form symmetry. When $\ell$ vanishes, they are responsible for the structure of charges as extended objects, consisting of the integral hypersurfaces of the $(p{-}1)$-form charge density $\rho^{i_2 ... i_p} \equiv \mathcal{O}^{t i_2 ... i_p}$. When $\ell \neq 0$, the integral hypersurfaces of the defect density $\tilde{\rho}^{i_3 ... i_p} \equiv \tilde{\mathcal{O}}^{t i_3 ... i_p}$ correspond to imperfections of the extended charges, that propagate in time along a defect --- as previously defined. The divergence of \cref{temporal} implies that the imperfections, as extended charges of a $(p{-}2)$-form symmetry, do not themselves possess imperfections. \\ \hspace*{1.75ex}
The spatial components \eqref{0.10}, on the other hand, are not constraints on a Cauchy surface but a set of dynamical equations. When $\ell = 0$, they mean that charges are conserved in time, as usual. In a hydrodynamical setting, they become equations of motion upon identifying the thermal expectation value of every \textit{flux density} $\braket{\mathcal{O}^{i_1 ... i_p}}$ with an expansion in gradients of $\braket{\rho_a} \equiv \braket{\rho_{i_2 ... i_p}}$ --- these are the constitutive relations that we denote by $\mathcal{J}_a^{i_1} \left( \partial_\mu^{n \geq 0} \braket{\rho_b} \right)$. When background sources are turned off, the hydrodynamical equations can be written as
\begin{equation} \begin{aligned} \label{Hydro_EOM}
	\partial_t \braket{\rho_a} + \partial_i \mathcal{J}_a^i \left( \partial_j^{n \geq 0} \braket{\rho_b} \right) = 0 \, .
\end{aligned} \end{equation}
Note that we have eliminated the dependence of $\mathcal{J}_a^i$ on the time derivatives of $\braket{\rho}$ by using the equation of motion itself. \\ \hspace*{1.75ex}
Finally, when $\ell$ is nonzero, \eqref{0.10} implies that charges are no longer conserved in time where $\tilde{\mathcal{O}}^{i_2 ... i_p} (x) \neq 0$. Hence, the $(d{-}1)$-dimensional integral hypersurfaces of the defect's flux density are not defects but junctions/boundaries that lie within a codimension-1 spacelike hypersurface. 
The question now is how does this case and, in particular, \cref{0.10} with $\ell \neq 0$ fit into the hydrodynamical framework. A possible answer, since $\ell$ is arbitrarily small, lies in \textit{extending} hydrodynamics. Note that, especially since the defect current itself is conserved, it is only natural to let the defect's flux density be given by a hydrodynamic constitutive relation. However, because both currents are coupled to each other, all constitutive relations must be an expansion in gradients of $\braket{\rho}$ and $\braket{\tilde{\rho}}$. In general, defects are not directly responsible for the non-conservation of charge, but indirectly via their fluxes.

\paragraph{}The present work aims to study effective holographic descriptions of systems with exact and approximate (continuous) higher-form symmetries. We focus on the probe limit of theories governing the low-energy dynamics of systems with a single higher-form symmetry. In the exact case, we consider bulk Maxwell-type theories, which capture a broad class of models found in the literature --- including those of \cite{Grozdanov:2017kyl, Grozdanov:2018ewh,Armas:2019sbe} and, in part, \cite{Hofman:2017vwr}. Explicit breaking, on the other hand, is realised following \cite{Pinheiro:2026ihi} via Proca-type theories. The dual field theories we consider, defined on the conformal boundary of AdS, are deformed by double-trace operators \cite{Aharony:2001pa}, with the deformation strength controlled by a parameter in the bulk theory. 
We derive their low-energy spectra at finite temperature by computing thermal (2-point) correlators of 
\begin{itemize}
	\item exactly and approximately conserved currents arising from electric symmetries; 
	\item Goldstones and multivalued Goldstones associated with magnetic symmetries.
\end{itemize}
We find that capturing the low-energy behaviour generally requires extending hydrodynamics into a more general effective field theory \cite{Boon&Yip} --- recently termed \textit{Hydro+} \cite{Stephanov:2017ghc} and \textit{Quasihydrodynamics} \cite{Grozdanov:2018fic} in slightly different contexts --- which we review in \cref{Quasihydro_sec}. This is true even when the higher-form symmetry is exact, provided the deformations are strong.

\paragraph{}The structure of the rest of the paper is as follows. In \cref{p2setup}, we lay out the family of holographic models studied in this work. Their defining properties are presented, as well as the electric-magnetic-like dualities relating them. In \cref{HForms_in_BH}, we first examine how ingoing boundary conditions are to be imposed on solutions to the equations of motion and then solve them perturbatively in frequency, wavenumber and, where applicable, mass. This constitutes the main computational effort in this work. In \cref{holog_quasihydro}, after reviewing quasihydrodynamics, we investigate how it describes the systems dual to our holographic models in the appropriate finite-temperature regimes. In \cref{correlator_results}, the results of \cref{HForms_in_BH} are used to compute retarded correlators. Besides confirming the analysis of \cref{holog_quasihydro}, this also supports the validity of a set of dualities that includes those mentioned above. We conclude in \cref{discussion} with a detailed summary of results and an outlook.

\paragraph{Conventions.} Coordinate indices on the $d$-dimensional physical spacetime with metric $\gamma$ (eventually taken to be the Minkowski metric $\eta$) are denoted by lowercase Greek letters $\mu, \nu, \ldots$. Among these, lowercase Latin letters $i, j, \ldots$ refer specifically to spatial coordinates. {We also decompose $x^i$ into the direction parallel to the wavevector ($x$) and the transverse directions ($x^A = y , z , \ldots$).} Lowercase Latin letters $a, b, \ldots$ from the beginning of the alphabet are used for indices in the $(d{+}1)$-dimensional bulk spacetime with metric $g$, whose boundary is identified with the physical spacetime. Greek and latin indices are raised with the inverse metrics $\gamma^{\mu \nu}$ and $g^{ab}$, respectively. Lastly, antisymmetrisation of indices is denoted with square brackets and it is not normalised, e.g. $X_{[a b]} = X_{a b} - X_{b a}$.

\section{Holographic setup} \label{p2setup}

\paragraph{}We are interested in the infrared dynamics characterised by continuous higher-symmetries of strongly interacting systems with holographic duals. According to the Gauge-gravity duality, the latter consist of gravity coupled to other fields $\Psi^{\text{(dyn)}}$,
\begin{equation} \begin{aligned}
	S = S_{\text{grav}} [g^{\text{(dyn)}}] + S_{\text{matter}} [g^{\text{(dyn)}} , \Psi^{\text{(dyn)}}] \, ,
\end{aligned} \end{equation}
in an asymptotically locally Anti de-Sitter (AlAdS) bulk spacetime. The dynamical metric $g^{\text{(dyn)}}$ encodes the gravitational degrees of freedom. \\ \hspace*{1.75ex}
In order to consider the system in a finite-temperature isotropic equilibrium state, we fix the bulk manifold $\mathbb{B}$ as an AlAdS$_{d+1}$ black brane with a background metric $g$ given by the line element
\begin{equation} \begin{aligned} \label{6.1}
	\mathrm{d} s^2 = {\mathrm{d} r^2 \over r^2 f (r)} - r^2 f (r) \mathrm{d} t^2 + r^2 \delta_{ij} \mathrm{d} x^i \mathrm{d} x^j .
\end{aligned} \end{equation}
The generic emblackening factor $f (r)$ is analytic at both the horizon $r = r_h$ and the conformal boundary. In particular, $f(r) = f' (r_h) (r-r_h) + \ldots$ and $f(r) = 1 + \ldots$  for ${r - r_h \over r_h} \ll 1$ and ${1 \over r} \ll 1$, respectively. The (Hawking) temperature $T$ is given by $4 \pi T = r_h^2 f' (r_h)$. \\ \hspace*{1.75ex}
Through Gauge-gravity duality, perturbations of the metric capture the behaviour of the boundary stress tensor. For the models we consider below, the remaining fields encode the dynamics of exactly or approximately conserved currents. In this paper, we neglect the backreaction of these fields on the bulk geometry (hence, we omit $S_{\text{grav}}$ from the actions below). Such a probe limit is relevant when the dynamics of the stress tensor and currents decouple from each other. This is e.g. the case for the low-density limit of higher-form charges in the thermal state.

\paragraph{}We first consider the case of an exact symmetry. Hence, we are interested in the dynamics of a conserved $p$-form current $\mathcal{O}$ such that $\mathrm{d} \ast \mathcal{O} = 0$. Holographically, this situation is dual to a $p$-form Maxwell field $A$, governed by the action
\begin{equation} \begin{aligned} \label{actionmassless_p2}
	\bar{S}_{\text{ren}} = {1 \over 2} \int_{\mathbb{B}} \mathrm{d}^{d+1} x {\sqrt{|g|} F_{a_0 ... a_p} F^{a_0 ... a_p} \over p+1} + \int_{\partial \mathbb{B}} \mathrm{d}^d x \, \mathscr{L}_{\text{counterterm}} [A] \, ,
\end{aligned} \end{equation}
where $F = \mathrm{d} A$. Holographic renormalisation via an appropriate choice of boundary Lagrangian is assumed to have been performed. On-shell, $A$ is the sum of two linearly independent solutions: one parametrised by $\alpha \in \Omega^{p}$, which is defined up to closed $p$-forms, and the other by $J \in \Omega^{p}$, which is co-closed. At large $r$, the leading on-shell behaviour is given by 
\begin{equation} \begin{aligned} \label{massless_soln}
	p! A_{\mu_1 ... \mu_p} & = \alpha_{\mu_1 ... \mu_{p}} + \ldots + {r^{1 - {\bar{\lambda}}} \over 1 - {\bar{\lambda}}} J_{\mu_1 ... \mu_p} + \ldots 
\end{aligned} \end{equation}
where ${\bar{\lambda}} \equiv d - 2 p+ 1$. When ${\bar{\lambda}}=1$, the coefficient of $J_{\mu_1 ... \mu_p}$ is $\ln r$ instead. The equation above (which assumes radial gauge $A_{r \mu_2 ... \mu_p} = 0$) can be seen as defining $\alpha$ and $J$ in terms of solutions $A$. The on-shell variation of the action is
\begin{equation} \begin{aligned} 
	\delta \bar{S}_{\text{ren}} = \int_{\partial \mathbb{B}} {\ast J \wedge \delta \alpha \over (d-p)!} \, .
\end{aligned} \end{equation} 
The integrand above agrees with the $r$-constant term in the on-shell variation of the bulk Lagrangian. From this one can see that we chose the boundary Lagrangian in \eqref{actionmassless_p2} as a minimal counterterm. We now consider that a term $\mathscr{L}_{\text{deformation}} [A]$ is added to $\mathscr{L}_{\text{counterterm}} [A]$. This term obeys $\mathscr{L}_{\text{deformation}} \vert_{\partial \mathbb{B}} \propto \mathcal{M} (\ast J \wedge J)$ on-shell --- with $\mathcal{M}$ as a coupling constant --- such that $\bar{S}_{\text{ren}}$ becomes $\bar{S}_{\text{final}}$, whose on-shell variation is given by
\begin{equation} \begin{aligned} 
	\delta \bar{S}_{\text{final}} = \int_{\partial \mathbb{B}} {\ast J \wedge \delta a \over (d-p)!} \, ,  \qquad \text{where} \qquad a \vcentcolon = \alpha + \mathcal{M} J \, .
\end{aligned} \end{equation} 
We introduce the term \textit{final variables}, which in this case consist of $J$ and $a$. \\ \hspace*{1.75ex}
With the bulk model settled, it remains only to present the holographic dictionary connecting it to the boundary theory where $\mathcal{O}$ lives. Before we do that, let us consider the case where the higher-form symmetry is explicitly broken such that $\mathcal{O}$ is no longer conserved. 
Based on \cite{Pinheiro:2026ihi}, we use a $p$-form $\mathcal{F}$ with mass $m^2 \neq 0$ in the bulk to account for the symmetry breaking. Its action is
\begin{equation} \begin{aligned} \label{action_p2}
	S_{\text{ren}} = {1 \over 2} \int_{\mathbb{B}} \mathrm{d}^{d+1} x \sqrt{|g|} \left[ {\mathcal{H}_{a_0 ... a_p} \mathcal{H}^{a_0 ... a_p} \over p+1} + m^2 \mathcal{F}_{a_1 ... a_p} \mathcal{F}^{a_1 ... a_p} \right] + \int_{\partial \mathbb{B}} \mathrm{d}^d x \, \mathscr{L}_{\text{counterterms}} [\mathcal{F}] \, ,
\end{aligned} \end{equation}
where $\mathcal{H} = (\mathrm{d} \mathcal{F} / p!)$. On-shell, we have
\begin{equation} \begin{aligned} 
	\mathcal{F}_{\mu_1 ... \mu_p} = r^{-\varDelta_-} K^{-}_{\mu_1 ... \mu_p} + \ldots + r^{-\varDelta_+} K^{+}_{\mu_1 ... \mu_p} + \ldots
\end{aligned} \end{equation}
at leading order for large $r$, where
\begin{equation} \begin{aligned} 
	\varDelta_\pm \equiv {\lambda - 3 \pm \sqrt{(\lambda - 3)^2 + 4 m^2} \over 2} \qquad \text{and} \qquad \lambda \equiv d - 2 p + 3 \, .
\end{aligned} \end{equation}
This component of $\mathcal{F}$ alone is sufficient to define the fields $K^- , K^+ \in \Omega^p$ which parametrise the solutions to the equations of motion. Note that the values $m^2 = - (\lambda-3)^2 / 4$, for which $\varDelta_+ = \varDelta_-$, are ruled out from our analysis since we are ultimately concerned with $|m^2| \ll 1$.
Unlike the massless case, this time there is not one but two minimal counterterms.\footnote{There is actually a one-parameter family of counterterms cancelling $r \to \infty$ divergences. However, only two elements in this family lead to a well-defined variational principle \cite{Pinheiro:2026ihi}.} They are inequivalent as they lead to $S_{\text{ren},-}$ and $S_{\text{ren},+}$ such that
\begin{subequations} \begin{align} 
	\delta S_{\text{ren},-} & = (\varDelta_+ - \varDelta_-) \int_{\partial \mathbb{B}} {\ast K^{-} \wedge \delta K^{+} \over (d-p)!} \\ 
	\delta S_{\text{ren},+} & = (\varDelta_- - \varDelta_+) \int_{\partial \mathbb{B}} {\ast K^{+} \wedge \delta K^{-} \over (d-p)!} \, .
\end{align} \end{subequations}
As before, we include a term $\mathscr{L}_{\text{deformation}} [\mathcal{F}]$ in the boundary lagrangian. On-shell,
\begin{equation} \begin{aligned}
	\mathscr{L}_{\text{deformation}} \vert_{\partial \mathbb{B}} \propto \mathcal{M}_{-} (\ast K^{-} \wedge K^{-}) \qquad \text{or} \qquad \mathscr{L}_{\text{deformation}} \vert_{\partial \mathbb{B}} \propto \mathcal{M}_{+} (\ast K^{+} \wedge K^{+}) \, , 
\end{aligned} \end{equation}	
such that $S_{\text{ren},-}$ and $S_{\text{ren},+}$ (divided by $\varDelta_+ - \varDelta_-$ and $\varDelta_- - \varDelta_+$) respectively become $S_{\text{final},-}$ and $S_{\text{final},+}$, whose on-shell variations are given by
\begin{subequations} \label{4.28_p2} \begin{align} 
	\delta S_{\text{final},-} & = \int_{\partial \mathbb{B}} {\ast K^{-} \wedge \delta \mathcal{K}^{+} \over (d-p)!} \, ,  \qquad \text{where} \qquad \mathcal{K}^+ \vcentcolon = K^{+} - \mathcal{M}_{-} K^{-} \\
	\delta S_{\text{final},+} & = \int_{\partial \mathbb{B}} {\ast K^{+} \wedge \delta \mathcal{K}^{-} \over (d-p)!} \, ,  \qquad \text{where} \qquad \mathcal{K}^- \vcentcolon = K^{-} - \mathcal{M}_{+} K^{+} \, .
\end{align} \end{subequations}

\paragraph{}It is now finally the time to give a precise relation between bulk and boundary theories. Boundary expectation values can be derived from a bulk path integral $Z$. In particular, we compute correlators of $\mathcal{O}$ in states with background source $\hat{\psi}$ by differentiating the generating functional
\begin{equation} \begin{aligned} \label{genfuncW0_p2}
	\expval{\exp( i \int_{\partial \mathbb{B}} {\ast \mathcal{O} \wedge (\psi - \hat{\psi}) \over (d-p)!})} = Z (\psi)
\end{aligned} \end{equation}
at $\psi = \hat{\psi}$. When $\mathrm{d} \ast \mathcal{O} = 0$, the path integral we use is
\begin{equation} \begin{aligned} \label{Za_p2} 
	\bar{Z}^{[{\bar{\lambda}} , \mathcal{M}]} \vcentcolon = \int_{a = \psi} \mathfrak{D} A e^{i \bar{S}_{\text{final}}}
\end{aligned} \end{equation}
and when $\mathrm{d} \ast \mathcal{O} \neq 0$, we use
\begin{equation} \begin{aligned} \label{Zpm_p2}
	Z^{[\lambda , \mathcal{M}_\mp]} \vcentcolon = \int_{\mathcal{K}^\pm = \psi} \mathfrak{D} \mathcal{F} e^{i S_{\text{final},\mp}} \, ,
\end{aligned} \end{equation}
where we have adopted the convention according to which the labels $\pm$ and $\mp$ are to be read respectively as $+$ and $-$, when $\lambda < 3$, or as $-$ and $+$, when $\lambda > 3$. (The case of $\lambda = 3$ is set aside). We choose to omit the coupling constant controlling quantum fluctuations of the path integrals in this section. Our focus in this work lies in the classical regime, holographically dual to the large-$N$ limit of the boundary theory.\footnote{\label{integ_domain}In the bulk path integrals, configurations that fail to satisfy the equations of motion at the boundary and regularity conditions in the interior are implicitly excluded from the integration domain.} \\ \hspace*{1.75ex}
The path integral \eqref{Za_p2} corresponds to an intact higher-form symmetry, while \eqref{Zpm_p2} leads to an explicit symmetry breaking controlled by the mass $m^2$. In particular, the symmetry is approximate when $|m^2| \ll 1$. 
These theories correspond to the \textit{electric quantisation} \cite{Pinheiro:2026ihi} of the massless and massive $p$-forms governed by the actions \eqref{actionmassless_p2} and \eqref{action_p2}, respectively. This is equivalent to standard quantisation when $d > 2 p$ and alternative when $d \leq 2 p$.

\paragraph{}The surface terms parametrised by $\mathcal{M}$ and $\mathcal{M}_\mp$ that we have considered are dual to double-trace deformations $\ast \mathcal{O} \wedge \mathcal{O}$ in the boundary theory. Because of them, boundary conditions on the bulk fields are in general not Dirichlet but Robin. In particular they fix $a$ and $\mathcal{K}^{\pm}$ as the boundary source --- cf. \cref{Za_p2,Zpm_p2}. \\ \hspace*{1.75ex}
The deformation with coupling constant $\mathcal{M}$ is relevant, marginal or irrelevant depending on ${\bar{\lambda}}$ being respectively lesser, equal or larger than 1. On the other hand, the deformations parametrised by $\mathcal{M}_{-}$ ($\mathcal{M}_{+}$) are relevant (irrelevant), such that $\mathcal{M}_\mp$ corresponds to a relevant or irrelevant deformation depending on $\lambda$ being respectively lesser or larger than 3. Note that ${\bar{\lambda}} = 1$ and $\lambda = 3$ both correspond to $p= d/2$.

\paragraph{}The construction above is relevant when we have an \textit{electric} $(p{-}1)$-form symmetry. If the boundary operator is a $p$-form gauge field $\mathcal{O} \sim \mathcal{O} + \mathrm{d} \zeta$, the field strength $\mathsf{f} \equiv \mathrm{d} \mathcal{O}$ is a local observable and $\ast \mathsf{f}$ is topologically conserved. We say in this case that the associated $(d{-}p{-}2)$-form symmetry is \textit{magnetic}. To distinguish between an (electric) current and a gauge-redundant operator, respectively, we label $\mathcal{O}$ according to $\mathcal{O}_j$ and $\mathcal{O}^a$. Because $\mathcal{O}_j$ is conserved, the source $\psi$ entering equation \eqref{genfuncW0_p2} is a background gauge field. That equation is also valid for $\mathcal{O}^a$, but with $\psi$ as a conserved background current. \\ \hspace*{1.75ex}
The holographic dual to $\mathcal{O}^a$ is still the Maxwell field $A$, but the bulk path integral from which we compute its correlators is no longer \eqref{Za_p2}. Consider the renormalised action \eqref{actionmassless_p2}. This time, we first add to $\mathscr{L}_{\text{counterterm}} [A]$ a term enforcing a Legendre transformation $\ast J \wedge \delta \alpha \to \ast \alpha \wedge \delta J$, and then we add a double-trace deformation as before. Hence, $\bar{S}_{\text{ren}}$ becomes $\bar{S}'_{\text{final}}$, whose on-shell variation is given by
\begin{equation} \begin{aligned} 
	\delta \bar{S}'_{\text{final}} = - \int_{\partial \mathbb{B}} {\ast \alpha \wedge \delta j \over (d-p)!} \, ,  \qquad \text{where} \qquad j \vcentcolon = J - \mathcal{M}^\prime {\ast \mathrm{d} \ast} \mathrm{d} \alpha \, .
\end{aligned} \end{equation} 
Note that this time the deformation comes from a term $\mathscr{L}_{\text{deformation}} [A]$ such that, on-shell, $\mathscr{L}_{\text{deformation}} \vert_{\partial \mathbb{B}} \propto \mathcal{M}^\prime (\ast \mathrm{d} \alpha \wedge \mathrm{d} \alpha)$. \\ \hspace*{1.75ex}
Lastly, we briefly address the case where the magnetic symmetry is explicitly broken. In this case, $\mathcal{O}^a$ no longer possesses the aforementioned gauge redundancy and, accordingly, the background current $\psi$ is no longer conserved. Before breaking the symmetry, we then use the path integral
\begin{equation} \begin{aligned} \label{Zj_p2} 
	\bar{Z}^{[{\bar{\lambda}} , \mathcal{M}^\prime]} \vcentcolon = \int_{j = \psi} \mathfrak{D} A e^{- i \bar{S}'_{\text{final}}} ,
\end{aligned} \end{equation}
while afterwards we use
\begin{equation} \begin{aligned} \label{Zmp}
	Z^{[\lambda , \mathcal{M}^\ast]} \vcentcolon = \int_{\mathcal{K}^\mp = \psi} \mathfrak{D} \mathcal{F} e^{i S_{\text{final},\pm}} 
\end{aligned} \end{equation}
with $\mathcal{M}^\ast \vcentcolon = \mathcal{M}_\pm = 0$. Turning on the respective deformation changes the symmetry that undergoes the explicit breaking (see \cite{Pinheiro:2026ihi} for further details). Nevertheless, we leave $\mathcal{M}^\ast$ unfixed throughout the paper. \\ \hspace*{1.75ex}
The bulk path integrals above correspond to the \textit{magnetic quantisation} \cite{Pinheiro:2026ihi} of the massless and massive $p$-forms governed by the actions \eqref{actionmassless_p2} and \eqref{action_p2}, respectively. This is equivalent to standard quantisation when $d \leq 2 p$ and alternative when $d > 2 p$. The degree of explicit symmetry breaking in this case is only proportional to the magnitude of $m^2$ when $\mathcal{M}^\ast = 0$, otherwise it is controlled by ${\varDelta_\pm \over \mathcal{M}^\ast} \sim {m^2 + O (m^4) \over \mathcal{M}^\ast}$. \\ \hspace*{1.75ex}
The deformation in \eqref{Zmp} has the same form $\ast \mathcal{O} \wedge \mathcal{O}$ as before. However, because when the magnetic symmetry is exact a deformation of this type is not gauge-invariant, in \eqref{Zj_p2} we considered a double-trace deformation $\mathcal{M}^\prime (\ast \mathrm{d} \mathcal{O}^a \wedge \mathrm{d} \mathcal{O}^a)$.\footnote{Note that, to emphasise the non-identical nature of the deformations used in magnetic quantisation, we assign different superscripts for the corresponding coupling constants.} This one is relevant, marginal or irrelevant depending on ${\bar{\lambda}}$ being respectively larger, equal or lesser than 3, while the one parametrised by $\mathcal{M}^\ast \equiv \mathcal{M}_\pm$ is a relevant or irrelevant deformation depending on $\lambda$ being respectively larger or lesser than 3. In the massive case, having introduced $\mathcal{M}^\ast$ and suppressed the $\pm$ subscript, we likewise omit the subscript in $\mathcal{M}_\mp$.

\paragraph{}The higher-form generalisation of electric-magnetic duality \cite{Misner:1957mt,Deser:1976iy} acts as a reflection ${\bar{\lambda}} \to 4 - {\bar{\lambda}}$ around ${\bar{\lambda}} = 2$, relating the equations of motion associated with the massless action \eqref{actionmassless_p2}. Using this duality, one can derive solutions for ${\bar{\lambda}} = {\bar{\lambda}}'$ from solutions for ${\bar{\lambda}} = 4 - {\bar{\lambda}}'$:
\begin{equation} \begin{aligned} \label{Hodgemap1_p2} 
	F^{(4-{\bar{\lambda}}')} \to F^{({\bar{\lambda}}')} & = u \star F^{(4-{\bar{\lambda}}')} ,
\end{aligned} \end{equation} 
where $u \in \mathbb{R}$. For the purposes of this paper, $u$ can be set to 1 without loss of generality. The solution $F^{({\bar{\lambda}}')}$ is then parametrised by a pair of $({d+1-{\bar{\lambda}}' \over 2})$-forms $\alpha^{({\bar{\lambda}}')}$ and $J^{({\bar{\lambda}}')}$ such that
\begin{subequations} \label{1.38_p2} \begin{align} 
	\beta^{({\bar{\lambda}}')} & = (-1)^{d+3-{\bar{\lambda}}' \over 2} \ast J^{(4-{\bar{\lambda}}')} \\
	J^{({\bar{\lambda}}')} & = \ast \beta^{(4-{\bar{\lambda}}')} \, , 
\end{align} \end{subequations}
where $\beta \equiv (\mathrm{d} \alpha / p!)$. This `massless' duality has a massive counterpart that acts as a reflection $\lambda \to 6 - \lambda$ around $\lambda = 3$. As before, this allows one to derive solutions of the equations of motion associated with the massive action \eqref{action_p2} according to
\begin{equation} \begin{aligned} \label{Hodgemap2_firstorder}
	\mathcal{F}^{(6-\lambda')} \to \mathcal{F}^{(\lambda')} = {v \star \mathrm{d} \mathcal{F}^{(6-\lambda')} \over (d-p)!} ,
\end{aligned} \end{equation}
 where $v \in \mathbb{R}$ is also set to 1 in the following.\footnote{In first-order formalism, one should rather consider 
 	\begin{equation} \begin{aligned} \label{Hodgemap2_p2}
 		\left\lbrace \mathcal{F}^{(6-\lambda')} \to \mathcal{F}^{(\lambda')} = v \star \mathcal{H}^{(6-\lambda')} , \mathcal{H}^{(6-\lambda')} \to \mathcal{H}^{(\lambda')} = (-1)^p m^2 v \star \mathcal{F}^{(6-\lambda')} \right\rbrace .
 	\end{aligned} \end{equation}
 } The solution $\mathcal{F}^{(\lambda')}$ is then parametrised by a pair of $({d+3-\lambda' \over 2})$-forms $K^{(\lambda') -}$ and $K^{(\lambda') +}$ such that
 \begin{subequations} \label{5.18_p2} \begin{align} 
 	K^{(\lambda') -} & = (-1)^{d+3-\lambda' \over 2} \varDelta_+ (\lambda') \ast K^{(6-\lambda') -} \\
 	K^{(\lambda') +} & = (-1)^{d+3-\lambda' \over 2} \varDelta_- (\lambda') \ast K^{(6-\lambda') +} \, .
 \end{align} \end{subequations}
Provided that the regularity conditions (satisfied by the configurations over which we integrate) are covariant under \eqref{Hodgemap1_p2} and \eqref{Hodgemap2_firstorder}, these dualities manifest themselves holographically via the following equivalence \cite{Pinheiro:2026ihi}:\footnote{At the level of correlation functions computed from holographic generating functionals.}
 \begin{subequations} \begin{align} 
 	\label{HodgeZ_massless_p2} 
 	\bar{Z}^{[{\bar{\lambda}}_1 , \mathcal{M}]} \leftrightarrow \bar{Z}^{[{\bar{\lambda}}_2 , \mathcal{M}^\prime]} , \quad \qquad {\mathcal{M} \over \mathcal{M}^\prime} & = 1 \\[1ex]
 	Z^{[\lambda_1 , \mathcal{M}]} \leftrightarrow Z^{[\lambda_2 , \mathcal{M}^\ast]} , \; \qquad {\mathcal{M} \over \mathcal{M}^\ast} & = {\varDelta_\pm (\lambda_I) \over \varDelta_\mp (\lambda_I)} \, ,
 \end{align} \end{subequations}
 where ${\bar{\lambda}}_1 + {\bar{\lambda}}_2 = 4$, $\lambda_1 + \lambda_2 = 6$ and $I=1,2$.\footnote{$\mathcal{M}^\prime$ has been rescaled according to $\mathcal{M}^\prime \to {(-1)^{p(d+1-p)} \over p! (d-p)!} \mathcal{M}^\prime$. In components, the definition of $j$ then becomes $j_{\mu_1 ... \mu_p} \vcentcolon = J_{\mu_1 ... \mu_p} - {\mathcal{M}^\prime \over p!} \partial^{\mu_0} \partial_{[\mu_0} \alpha_{\mu_1 ... \mu_p]}$.} Note, in particular, that ${\varDelta_\pm \big/ \varDelta_\mp} = {- m^2 \big/ (\lambda-3)^2} + O (m^4)$. \\ \hspace*{1.75ex} 
A $p$-form operator with gauge redundancy $\mathcal{O}^a \sim \mathcal{O}^a + \mathrm{d} \zeta$ arises as the Goldstone field of a spontaneously broken (continuous) $p$-form symmetry and therefore constitutes a natural low-energy effective degree of freedom.
Our naming conventions assume that we use `the Goldstone formulation' as the starting point for effective holographic descriptions of systems with a magnetic symmetry. However, both electric and magnetic quantisation would {a priori} be valid and moreover, in the presence of electric-magnetic duality, they are dynamically equivalent. The possibility of two different formulations is useful in the context of hydrodynamics and has been used for example with viscoelasticity \cite{Armas:2019sbe}.

\paragraph{}In the massive case, there is a second duality at play (see e.g. \cite{Pinheiro:2026ihi} and references therein). This is a strong/weak-coupling duality --- on $\mathcal{M}$ and $\mathcal{M}^\ast$ --- according to which, under identical Robin boundary conditions, correlation functions in different quantisations differ only by contact terms. We conclude by remarking that the discussion and expressions presented in this section remain valid for any AlAdS metric in Fefferman-Graham $(r , x^\mu)$ coordinates (the transverse metric approaches $\gamma$ as $r \to \infty$).

\section{Higher-form fields in planar-horizon geometries} \label{HForms_in_BH}

\paragraph{}We consider the equations of motion associated with the actions \eqref{actionmassless_p2} and \eqref{action_p2} outside the horizon of the black brane \eqref{6.1}. We will be solving them in first-order formalism (cf. \cite{Pinheiro:2026ihi}) for the Fourier-transformed bulk fields: $F$, $\mathcal{F}$ and $\mathcal{H}$. To this end, we use the equivalence between $F_{a_0 ... a_p}$ and
\[ \int {\mathrm{d} \omega \mathrm{d} k \over (2 \pi)^2} F_{a_0 ... a_p} (r , \omega , k) e^{-i \omega t + i k x} , \]
with an analogous expression for $\mathcal{F}_{a_1 ... a_p}$ and $\mathcal{H}_{a_0 ... a_p}$. There are $d-2$ transverse coordinates such that $x^\mu \equiv (t, x, x^A)$ where $A = 2 , \ldots , d$. Without loss of generality, these directions were chosen (using rotational symmetry) so the wavevector vanishes along them, $k^A = 0$. \\ \hspace*{1.75ex}
Our convention for raising transverse indices follows naturally from our previous convention: we will use $\eta^{\mu \nu}$ to raise them in the boundary fields' components (which makes the up/down position of transverse indices irrelevant in this case) and for the bulk fields we use $g^{ab}$. Additionally, we adopt
\vspace*{0.4em}

\noindent \hypertarget{conv2}{\textbf{(convention 1):}} transverse indices are often omitted in all bulk and boundary fields' components except $F_{A_0 ... A_p}$ and $\mathcal{F}_{A_1 ... A_p}$. For example: $F_{t}$ stands for $F_{t A_1 ... A_p}$ and $F^{t}$ stands for $F^{t A_1 ... A_p}$, such that indices are either all up or all down. \hangindent=2em \hangafter=0

\vspace*{0.4em}  

\paragraph{}This section is organised as follows. We start by solving the equations of motion near the horizon in order to impose ingoing boundary conditions there. Then, in \cref{hydrolim}, we solve respectively the massless and massive equations perturbatively in gradients.

\subsection{Ingoing solutions at the horizon} \label{nearhorizon}

\paragraph{}We are solving the equations of motion in Fourier space and have aligned the momentum of each plane-wave such that $k^\mu = (\omega, k, 0, \ldots , 0)$. Because the equations are still explicitly covariant under rotations in the $x^A$-plane, they decouple into several closed subsystems, each of which corresponds to a different representation of $SO(d-2)$. As the equations are linear, each subsystem may be labelled by the number $p^\perp$ of transverse indices in a given component of $F$ or $\mathcal{F}$, or by
\begin{equation} \begin{aligned}
	{\bar{\lambda}}_{\text{eff}} \vcentcolon = 3 - 2 (p- p^\perp) \quad \quad \quad \text{and} \quad \quad \quad \lambda_{\text{eff}} \vcentcolon = 5 - 2 (p - p^\perp) \, ,
\end{aligned} \end{equation} 
equivalently.

\paragraph{}Starting with the massless equations, these include four subsystems, two of which are trivial in the sense that they simply set $F_{A_0 ... A_p}$ and $\sqrt{|g|} F^{r t x}$ to be constant with respect to $r$, $t$ and $x$. The first non-trivial system is
\begin{subequations} \label{2.2} 
	\begin{empheq}[left={{\bar{\lambda}}_{\text{eff}} = 3: \quad \quad}]{align}
		\label{2.2a} (\mathrm{d}^\dagger F)^{A_1 ... A_p} & = 0 \\
		\label{2.2b} (\mathrm{d} F)_{a b A_1 ... A_p} & = 0 \, , \quad \quad a , b \in \{ r , t , x \} .
	\end{empheq}
\end{subequations} 
This system is present when $0 \leq p\leq d-2$, whereas a second non-trivial one arises in the range $1 \leq p\leq d-1$ and is
\begin{subequations} \label{eq:2.3} 
	\begin{empheq}[left={{\bar{\lambda}}_{\text{eff}} = 1: \quad \quad}]{align}
		\label{eq:2.3a} (\mathrm{d}^\dagger F)^{a A_2 ... A_p} & = 0 \, , \quad \quad a \in \{ r , t , x \} \\
		\label{eq:2.3b} (\mathrm{d} F)_{r t x A_2 ... A_p} & = 0 \, .
	\end{empheq}
\end{subequations} 
Note that, under the Hodge map \eqref{Hodgemap1_p2}, the systems above are mapped to one another. \Cref{2.2a,eq:2.3b} can be respectively manipulated (by using the remaining equations in each system) into the following 2nd-order ODEs: 
\begin{subequations} \label{6.5} \begin{align} 
	\label{6.5a} 0 & = {f(r) \over r^{{\bar{\lambda}}-4}} \partial_{r} \left( r^{{\bar{\lambda}}} f(r) \partial_{r} F_{t} \right) - \left( \partial^2_{t} - f(r) \partial^2_{x} \right) F_{t} \\
	\label{6.5b} 0 & = r^{{\bar{\lambda}}} f (r) \partial_r \left( {f(r) \over r^{{\bar{\lambda}}-4}} \partial_{r} \left(  \sqrt{|g|} F^{r t} \right) \right) - \left( \partial^2_t - f(r) \partial^2_x \right) \sqrt{|g|} F^{r t} \, .
\end{align} \end{subequations}

\paragraph{}The massive equations also decouple into several closed subsystems. The trivial ones simply set $\mathcal{H}_{A_0 ... A_p}$ and $\sqrt{|g|} \mathcal{F}^{r t x}$ to be constant with respect to $r$, $t$ and $x$. There are three non-trivial subsystems, two of which are
\begin{subequations} \label{23}
	\begin{empheq}[left={\lambda_{\text{eff}} = 5: \quad \quad}]{align}
		\label{23a} (\mathrm{d}^\dagger \mathcal{H} - m^2 \mathcal{F})^{A_1 ... A_p} & = 0 \\
		\label{23b} (\mathrm{d} \mathcal{F} - p! \mathcal{H})_{a A_1 ... A_p} & = 0 \, , \quad \quad a \in \{ r , t , x \},
	\end{empheq}
\end{subequations} 
present when $0 \leq p \leq d-2$, and
\begin{subequations} \label{26}
	\begin{empheq}[left={\lambda_{\text{eff}} = 1: \quad \quad}]{align}
		\label{26a} (\mathrm{d}^\dagger \mathcal{H} - m^2 \mathcal{F})^{a b A_3 ... A_p} & = 0 \, , \quad \quad a , b \in \{ r , t , x \} \\
		\label{26b} (\mathrm{d} \mathcal{F} - p! \mathcal{H})_{r t x A_3 ... A_p} & = 0 \, , 
	\end{empheq}
\end{subequations} 
present when $2 \leq p \leq d$. Under the Hodge map \eqref{Hodgemap2_p2}, the systems above are mapped to one another. \Cref{23a,26b} can be respectively manipulated into: 
\begin{subequations} \label{6.8} \begin{align} 
	\label{6.8a} 0 & = r^{6-\lambda} \partial_{r} \left( f(r) r^{\lambda-2} \partial_{r} \mathcal{F}_{A_1 ... A_p} \right) - \left( {\partial_{t}^2 \over f (r)} - \partial_{x}^2 + m^2 r^2 \right) \mathcal{F}_{A_1 ... A_p} \\
	\label{6.8b} 0 & = r^{\lambda} \partial_{r} \left[ f (r) r^{4-\lambda} \partial_{r} \left( \sqrt{|g|} \mathcal{H}^{r t x} \right) \right] - \left( {\partial_{t}^2 \over f (r)} - \partial_{x}^2 + m^2 r^2 \right) \sqrt{|g|} \mathcal{H}^{r t x} \, .
\end{align} \end{subequations}

\paragraph{}The last non-trivial subsystem exists for $1 \leq p \leq d - 1$ and is
\begin{subequations} \label{24}
	\begin{empheq}[left={\lambda_{\text{eff}} = 3: \quad \quad}]{align}
		\label{24a} (\mathrm{d}^\dagger \mathcal{H} - m^2 \mathcal{F})^{a A_2 ... A_p} & = 0 \\
		\label{24b} (\mathrm{d} \mathcal{F} - p! \mathcal{H})_{b c A_2 ... A_p} & = 0 \, , \quad \quad a , b , c \in \{ r , t , x \} .
	\end{empheq}
\end{subequations} 
Unlike in the previous cases, the action of the Hodge map in this system is still an automorphism of its equations. In particular, equations \eqref{24a} are mapped to equations \eqref{24b} and vice-versa. 

\paragraph{}Let us denote the variables in \eqref{6.5} and \eqref{6.8} collectively by $Y$. Near the horizon, all four of these equations are of the form $r_h^4 f'(r_h)^2 \partial^2_{\rho} Y + \ldots = \partial_t^2 Y$, where $\rho = \ln [ f'(r_h) (r-r_h)]$. Its solutions in coordinates $(\rho, x^\mu)$ are the sum of an ingoing and an outgoing wave (relative to the horizon). Hence, introducing the $r$-constants $\Gamma[Y]$ and $\Xi[Y]$, the solutions to the equations in \eqref{6.5} and \eqref{6.8} have a near-horizon, low-frequency expansion of the form
\begin{equation} \begin{aligned} 
	Y(r \approx r_h) & = \ln f \left( \Xi[Y] + \ldots \right) + \left( \Gamma[Y] + \ldots \right) + O \left( {\omega \over T} \right)^2 .
\end{aligned} \end{equation} 
Since our ultimate goal is to compute retarded correlators in the boundary theory, we use ingoing-wave boundary conditions at the horizon \cite{Son:2002sd,Herzog:2002pc}. This imposes
\begin{equation} \begin{aligned} \label{6.13}
	\Xi (Y) = {\partial_t \Gamma (Y) \over 4 \pi T} \, ,
\end{aligned} \end{equation}
for $Y \in \{ F_{t} , \sqrt{|g|} F^{r t} , \mathcal{F}_{A_1 ... A_p} , \sqrt{|g|} \mathcal{H}^{r t x} \}$. For the $\lambda_{\text{eff}} = 3$ system we find, not one PDE like $r_h^4 f'(r_h)^2 \partial^2_{\rho} Y + \ldots = \partial_t^2 Y$, but a pair of coupled equations. Still, these are solved (near the horizon) by ingoing waves such that \eqref{6.13} can be extended to $Y \in \{ \sqrt{|g|} \mathcal{F}^{r} , \mathcal{F}_{x} , \mathcal{F}_{t} \}$.\footnote{The ingoing boundary conditions in terms of  the components of $\mathcal{H}$ governed by the $\lambda_{\text{eff}} = 3$ system will not be necessary.} The details regarding the $\lambda_{\text{eff}} = 3$ system can be found in appendix \ref{appendD}.

\subsection{Ingoing solutions outside the horizon} \label{hydrolim}

\paragraph{}We start with the massless equations, i.e. the systems corresponding to ${\bar{\lambda}}_{\text{eff}} = 3,1$. Our goal is to express the \textit{ingoing wave conditions} \eqref{6.13} in terms of the $r$-constants that parametrise the near-boundary solution \eqref{massless_soln}. We relegate most of the technical details to appendix \ref{appendD} and restrict the presentation here to illustrating the overall structure of the procedure by explicitly solving the case of ${\bar{\lambda}}_{\text{eff}} = 3$.

\paragraph{}Instead of \cref{6.13} for $Y = F_{t}$, we use
\begin{equation} \begin{aligned} \label{6.6b} 
	\Xi (\sqrt{|g|} F^{r}) & = {\partial_t \Gamma (\sqrt{|g|} F^{r}) \over 4 \pi T} \, ,
\end{aligned} \end{equation}
which is equivalent by virtue of the equations in \eqref{2.2}. Our goal is then to express this relation in terms of $J \in \Omega^p$ and $\beta \in \Omega^{p+1}$, by substituting $\sqrt{|g|} F^{r}$ on-shell. We integrate \cref{2.2a} and the radial components of \cref{2.2b} and obtain
\begin{subequations} \label{9.14} \begin{align} 
	\label{9.14a} \sqrt{|g|} F^{r} & = J + \partial_{t} \int \mathrm{d} r {r^{{\bar{\lambda}}-4} \over f (r)} F_{t} - \partial_{x} \int \mathrm{d} r r^{{\bar{\lambda}}-4} F_{x} \\
	\label{9.14b} F_{\mu} & = \beta_{\mu} + \partial_{\mu} \int \mathrm{d} r {\sqrt{|g|} F^{r} \over r^{\bar{\lambda}} f (r)} \, ,
\end{align} \end{subequations}
where $\mu \in \{ t , x \}$ and we identified integration constants with boundary fields by comparison with \eqref{massless_soln}. Note that indefinite integrals are defined up to a constant which is always chosen so the integral has no $r$-independent term when expanded as $r \to \infty$. \\ \hspace*{1.75ex} 
Inserting \cref{9.14b} in \eqref{9.14a}, we arrive at
\begin{equation} \begin{aligned} \label{6.9}
	\sqrt{|g|} F^{r} & = J + \partial_{t} \beta_{t} \int \mathrm{d} r {r^{{\bar{\lambda}}-4} \over f (r)} - \partial_{x} \beta_{x} {r^{{\bar{\lambda}}-3} \over {\bar{\lambda}}-3} + O (\omega^2 , k^2) F^r \, .
\end{aligned} \end{equation}
In conclusion,\footnote{The coefficient of $\ln f (r)$ in the near-horizon expansion of \eqref{6.9} can be identified using the formula \eqref{formula} for $h (r) = r^{{\bar{\lambda}}-4}$.} \cref{6.6b} is equivalent to
\begin{equation} \begin{aligned} \label{7.4}	
	\beta_{t} r_h^{{\bar{\lambda}}-2} & = J - \partial_{x} \beta_{x} {r_h^{{\bar{\lambda}}-3} \over {\bar{\lambda}}-3} + O (\omega , k^2) F^r .
\end{aligned} \end{equation}
Using the fact that $\beta_\mu = \partial_\mu \alpha$,\footnote{Hence, the non-radial component of \cref{2.2b} (which is the only radial constraint in that system) is identically satisfied.} this equation defines a map between $\alpha$ and $J$. Repeating a similar sequence of steps (cf. appendix \ref{appendD}), one obtains 
\begin{equation} \begin{aligned} \label{7.7}	
	J^{x} r_h^{2-{\bar{\lambda}}} & = \beta_{t x} + \partial_{x} J^{t} {r_h^{1-{\bar{\lambda}}} \over 1-{\bar{\lambda}}} + O (\omega , k^2) F_{t x} \, ,
\end{aligned} \end{equation}
for the ${\bar{\lambda}}_{\text{eff}} = 1$ system. The only massless equation we have yet to solve is the radial component of \cref{eq:2.3a}. This is simply the constraint which imposes the conservation equation $\partial_t J^t + \partial_x J^x = 0$. Since $\beta_{t x} = \partial_{[t} \alpha_{x]}$, this equation and the one above define a unique map between $J^{\mu = t , x}$ and $\alpha_{\mu = t , x}$. \\ \hspace*{1.75ex}
Anticipating \cref{holog_interp}, \cref{7.7} takes the form of a constitutive relation for $J^{x}$. Similarly, \cref{7.4} takes the form of a \textit{Josephson equation} 
governing the dynamics of a Goldstone field --- here, $\alpha$ --- in response to an external current.
In the present case, these two equations can be obtained from each other using \eqref{1.38_p2}, which simply reflects the fact that systems ${\bar{\lambda}}_{\text{eff}} = 3$ and ${\bar{\lambda}}_{\text{eff}} = 1$ are related via the Hodge map. Because the conservation equation is equally dual to the Bianchi identity $\partial_{t} \beta_{x} = \partial_{x} \beta_{t}$, the former (upon substituting the constitutive relation) is indeed equivalent to the Josephson equation \eqref{7.4}.\footnote{The terminology ‘Josephson equation’ originates from its use in superfluid hydrodynamics \cite{LandauLifshitz1959,Putterman1974}.} \\ \hspace*{1.75ex}
Before proceeding, note that we have been implicitly assuming that $3 \neq {\bar{\lambda}} \neq 1$. In order to lift this restriction, we introduce the following notation:
\begin{equation} \begin{aligned} \label{6.22}
	{1 \over {\bar{\lambda}} - 3} \equiv \ln r_h \quad \text{ when} \quad {\bar{\lambda}} = 3 \, ; \quad \quad {1 \over 1 - {\bar{\lambda}}} \equiv \ln r_h \quad \text{when} \quad {\bar{\lambda}} = 1 \, .
\end{aligned} \end{equation}

\paragraph{}Our next goal is to find the consequences of imposing ingoing boundary conditions in massive theories. Starting with the systems corresponding to $\lambda_{\text{eff}} = 5,1$, we find that the ingoing wave conditions \eqref{6.13} for $Y \in \{ \mathcal{F}_{A_1 ... A_p} , \sqrt{|g|} \mathcal{H}^{r t x} \}$ are equivalent to
\begin{subequations} \label{9.27} \begin{align} 
	\label{7.23} \partial_{t} K^{\pm} r_h^{\lambda-4} & = (3-\lambda) K^{\mp} - \partial_{x}^2 K^{\pm} {r_h^{\lambda-5} \over \lambda-5} + m^2 K^{\pm} {r_h^{\lambda-3} \over \lambda-3} + O ( \omega , k^2 , m^2) \mathcal{H}^{r} \\
	\label{7.26} \partial_{t} K^{\mp}_{t x} r_h^{2-\lambda} & = {m^2 \over 3-\lambda} K^{\pm}_{t x} - \partial_{x}^2 K^{\mp}_{t x} {r_h^{1-\lambda} \over 1-\lambda} + m^2 K^{\mp}_{t x} {r_h^{3-\lambda} \over 3-\lambda} + m^2 O (m^2 , \omega , k^2) \mathcal{F}_{t x} \, ,
\end{align} \end{subequations}
These follow from a derivation --- cf. appendix \ref{appendD} --- that parallels what we did above for the ${\bar{\lambda}}_{\text{eff}} = 3$ system. Note that we are using the convention from \cref{p2setup} regarding the $\pm$ and $\mp$ labels.
In the equations above, we have implicitly assumed that $5 \neq \lambda \neq 1$. To overcome this limitation, we introduce notation such that \eqref{7.23} and \eqref{7.26} hold for all $\lambda$:
\begin{equation} \begin{aligned}  
	{1 \over \lambda - 5} \equiv \ln r_h \quad \text{when} \quad \lambda = 5 \, ; \quad \quad {1 \over 1 - \lambda} \equiv \ln r_h \quad \text{when} \quad \lambda = 1 \, .
\end{aligned} \end{equation}

\paragraph{}Finally, for the $\lambda_{\text{eff}} = 3$ system, we introduce $X^{-} , X^{+} \in \Omega^{p-1}$ parametrising the on-shell near-boundary expansion of the radial components of $\mathcal{F}$. Of particular importance is $X^{\mp}$, which satisfies the approximate conservation equation (cf. \cite{Pinheiro:2026ihi}) that follows from the radial component of \eqref{24a}:
\begin{equation} \begin{aligned} \label{6.57} 
	\partial^\mu K^{\mp}_\mu = {m^2 + O (m^4) \over \lambda-3} X^{\mp} .
\end{aligned} \end{equation}
Details are once again collected in appendix \ref{appendD}, but we focus on the fact that the ingoing wave conditions \eqref{6.13} for $Y \in \{ \mathcal{F}_{x} , \mathcal{F}_{t} \}$ are equivalent to
\begin{subequations} \label{6.45} \begin{align} 
	r_h^{2-\lambda} X^{\mp} & = K^{\pm}_t + r_h^{3-\lambda} K^{\mp}_t + \partial_{x} K^{\pm}_x {r_h^{-1} \over \lambda-3} + \ldots \\
	r_h^{2-\lambda} \partial_{x} X^{\mp} & = \partial_t K^{\pm}_x + \partial_{x}^2 K^{\pm}_x {r_h^{-1} \over \lambda-3} + (\lambda-3) K^{\mp}_x r_h^{4-\lambda} - K^{\pm}_x {r_h m^2 \over \lambda-3} + \ldots 
\end{align} \end{subequations}
To avoid clutter, the order of subleading terms has exceptionally been omitted. Interpreting the equations above as constitutive relations for $X^{\mp}$ and $K_x^{\mp}$, \cref{6.57} would become a hydrodynamic equation if not for a linear term in the right-hand side. Equations of this type will play an important role in the next section. \\ \hspace*{1.75ex}
For the 2-point correlators computed in \cref{correlator_results}, \cref{6.57} will simply be used to eliminate $X^{\mp}$ from \eqref{6.45}. Let us introduce the dimensionless frequency and wavenumber such that $(\omega , k) = r_h (\hat{\omega} , \hat{k})$, the parameter $0 < \varepsilon \ll 1$ and consider $\hat{k} \sim \varepsilon \sim m$. In the scaling limit where $\hat{\omega} \sim \varepsilon$, we obtain
\begin{equation} \begin{aligned} \label{7.47}
	r_h^{\lambda-3} \begin{pmatrix}
		i {m^2 \over 3-\lambda} + O (\varepsilon^3) & \hat{k} {m^2 \over (\lambda-3)^2} + O (\varepsilon^4) \\
		O (\varepsilon^4) & i \hat{\omega} {m^2 \over \lambda-3} + O (\varepsilon^4)
	\end{pmatrix}
	\begin{pmatrix}
		K^{\pm}_t \\
		K^{\pm}_x
	\end{pmatrix} 
	=
	\begin{pmatrix}
		\hat{\omega} + O (\varepsilon^2) & \hat{k} + O (\varepsilon^2) \\
		\hat{\omega} \hat{k} + O (\varepsilon^3) & \hat{k}^2 + m^2 + O (\varepsilon^3)
	\end{pmatrix} \begin{pmatrix}
		K^{\mp}_t \\
		K^{\mp}_x
	\end{pmatrix} .
\end{aligned} \end{equation}
Note that, unlike in equations \eqref{6.45}, we are once again keeping track of the order of subleading terms. 

\section{Holographic quasihydrodynamics} \label{holog_quasihydro}

\paragraph{}In this section, we first introduce the aspects of quasihydrodynamics relevant to our purposes, before applying them to the holographic results of the previous section. Although the main results of this paper are the correlators computed in the following section, \cref{holog_interp} provides a preliminary analysis of the dynamics dual to our bulk models, particularly in the massless case. 
For this, our starting point is that: as is standard, the holographic dictionary implies the on-shell variation of the `final actions' from \cref{p2setup} to be such that, in the classical large-$N$ limit,
\begin{equation} \begin{aligned}
	\delta \mathcal{S}_{\text{final}} \sim \int_{\partial \mathbb{B}} \ast [\text{expectation value's dual}] \wedge \delta [\text{source's dual}] \, .
\end{aligned} \end{equation}
This allows us to identify \textit{final variables} introduced previously within the boundary theory.

\subsection{Quasihydrodynamics} \label{Quasihydro_sec}

\paragraph{}Hydrodynamics is an effective field theory (EFT) for many-body systems near global equilibrium at finite temperature \cite{Liu:2018kfw,Kovtun:2012rj}. 
In a top-down perspective on EFTs \cite{Polchinski:1992ed}, one starts with quantum fields that can be expressed as a sum of Fourier modes $u_k$, i.e. \[\varphi (x) = \sum \left( a_k u_k (x) + \mathrm{c.c.} \right) \qquad \text{where} \qquad \partial_t u_k = - i \omega u_k ,\] and splits them into low and high-frequency parts with respect to a cutoff $\varLambda$: \[\varphi_- (x) \vcentcolon = \sum_{\omega < \varLambda} \left( a_k u_k (x) + \mathrm{c.c.} \right) \qquad \text{and} \qquad \varphi_+ \vcentcolon = \varphi - \varphi_- .\] Hence, one can write the QFT path integral $\mathcal{Z} = \int \mathfrak{D} \varphi e^{i I [\varphi]}$ as \[\mathcal{Z} = \int \mathfrak{D} \varphi_- \left( \int \mathfrak{D} \varphi_+ e^{i I [\varphi_- , \varphi_+]} \right) \equiv \int \mathfrak{D} \varphi_- e^{i I_\varLambda [\varphi_-]} ,\]
where we have introduced the low-energy effective action $I_\varLambda$. Although the top-down viewpoint helps organise the conceptual framework, hydrodynamical theories are generally formulated from the bottom up. 
Hydrodynamics is particularly useful, when a perturbation of low wavenumber $k$ drives a system away from equilibrium, in describing the relaxation of the \textit{slow variables}. These are macroscopic fields relevant for the dynamics of the low-energy effective action and determined by the symmetries of the system --- we consider a set of locally conserved charge densities $\{\rho\}$ for concreteness but, e.g. in the presence of spontaneous symmetry breaking, they can be the phase of the relevant order parameter \cite{Lubensky,forster_1975} --- our so-called \textit{Goldstone field}. Slow variables have long \textit{relaxation times} such that 
\begin{equation} \begin{aligned} \label{7.2}
	\rho - \rho_{\text{equilibrium}} \sim \exp({-t \over \tau (k)}) \, , \quad \quad \tau (k) \gg \Delta t \, ,
\end{aligned} \end{equation}
where $\Delta t$ is a characteristic time scale of the system. This means that we expect the retarded Green's functions of $\rho$ (Fourier transformed) to have poles in the low-frequency, low-wavenumber region of the complex $\omega \,$--$\, k$ plane. According to hydrodynamics, the relaxation time becomes singular when $k$ goes to zero. In terms of dispersion relations of collective charge excitations, we have
\begin{equation} \begin{aligned}
	{1 \over \tau_{\text{hydro}} (k)} \equiv i \omega (k) \quad \quad \text{such that} \quad \quad \lim_{|k| \to 0} |\omega (k)| = 0 \, ,
\end{aligned} \end{equation}
reflecting the characteristic gaplessness of \textit{hydrodynamic modes}. The dispersion relations of hydrodynamic modes are given as low-$|k|$ Taylor expansions with a finite radius of convergence $k_\star$. One often fixes $\Delta t^{-1}$ as the microscopic scale associated with the breakdown of hydrodynamics as a valid EFT, in which case $\Delta t^{-1} \sim |\omega (k_\star)|$ --- this is found to be order $T$ in many instances of strongly interacting systems, such as the ones with holographic duals. Without loss of generality, we will then denote such a microscopic scale by $T$ in the following.

\paragraph{}Like in any EFT, it is useful to introduce a UV cutoff (with dimensions of energy). We consider the cutoff $\Lambda_{\text{Hydro}}$ such that $\tau_{\text{hydro}} (k)^{-1} < \Lambda_{\text{hydro}} \ll T$ and ask the following: \\[1ex] 
\hspace*{\fill} what happens if  we consider a new EFT ($\equiv$ EFT$_{\text{QH}}$) by raising it to $\Lambda_{\text{QH}} \ll T$? \hspace*{\fill} \\[1ex] 
In principle, this means the EFT$_{\text{QH}}$ can describe how the system responds to perturbations with larger wavenumber than before. We are interested in the scenario where the spectrum of the new theory contains additional modes with a parametrically small gap, such that their dispersion relations obey
\begin{equation} \begin{aligned}
	\Lambda_{\text{QH}} > |\Im \omega (k)| \gtrsim {1 \over \tau_{\text{gap}}} \, ,
\end{aligned} \end{equation}
where $\tau_{\text{gap}}$ is another time scale characteristic of the system. We will refer to these as quasihydrodynamic modes and to the EFT$_{\text{QH}}$ as quasihydrodynamics \cite{Grozdanov:2018fic}.
With the benefit of hindsight, we distinguish between the following cases:
\begin{enumerate}
	\item the quasihydrodynamic modes are carried by the locally conserved charge densities $\{\rho\}$;
	\item there are additional slow variables $\{ \mathfrak{p} \}$, such as charge densities $\tilde{\rho}$ of weakly broken symmetries,\footnote{Or even flux densities of currents associated with exact or approximate symmetries.} associated with quasihydrodynamic excitations.
\end{enumerate}
We start by addressing the second situation. Recall that classical hydrodynamics is given by a set of equations governing the thermal expectation values of $\{\rho\}$, which take the form of a conservation equation \eqref{Hydro_EOM} --- there, $\mathcal{J}$ denoted constitutive relations for the components of conserved currents corresponding to flux densities, which can be expressed as a gradient expansion of charge densities.\footnote{Transport coefficients can be seen as the Wilson coefficients of hydrodynamics.} We expect these equations to be modified by the fact that $\{\mathcal{J}\}$ can depend on the slow variables $\{ \mathfrak{p} \}$ besides $\{\rho\}$. Hence, the classical equations of motion of EFT$_{\text{QH}}$ include
\begin{equation} \begin{aligned} \label{hydroquasi} 
	\partial_t \braket{\rho} + \partial_i \mathcal{J}^i \left( \partial_j^{n \geq 0} \braket{\rho} , \partial_j^{n \geq 0} \braket{\mathfrak{p}} \right) & = 0 \, ,
\end{aligned} \end{equation}
where we take $\braket{\rho}$, $\braket{\mathfrak{p}}$ and $\mathcal{J}^i$ to be vectors. In the case where $\{ \mathfrak{p} \}$ are charge densities of weakly broken symmetries, the equations of motion also feature
\begin{equation} \begin{aligned} \label{quasihydro} 
	\partial_t \braket{\mathfrak{p}} + \partial_i \mathcal{P}^i \left( \partial_j^{n \geq 0} \braket{\rho} , \partial_j^{n \geq 0} \braket{\mathfrak{p}} \right) & = - (\tau_{\text{gap}})^{-1} \cdot \braket{\mathfrak{p}} \, ,
\end{aligned} \end{equation}
where $\mathcal{P}^i$ is a vector of constitutive relations for flux densities of approximately conserved currents. We assume that a basis of $\{ \mathfrak{p} \}$ has been chosen such that the matrix $(\tau_{\text{gap}})^{-1}$ is diagonal.

\paragraph{}Regarding point 1 above, such a situation is realised at the classical level when the constitutive relations $\{\mathcal{J}\}$ in \eqref{Hydro_EOM} cease to be local and thus cannot be expressed as a gradient expansion. Working in Fourier space, the coefficients of $\{ \braket{\rho} \}$ in the constitutive relations become non-analytic, with singularities located close to --- but off --- the origin of the $\omega \,$--$\, k$ plane. 
A particular example that will later be of interest to us is that of a diffusion-to-sound crossover \cite{Grozdanov:2018fic}, when we have a single charge density $\rho$ and flux $\mathcal{J}$ such that
\begin{equation} \begin{aligned} \label{eq:7.7}
	\mathcal{J} \approx {i k \braket{\rho} \over i \omega - \tau_{\text{gap}}^{-1}} \, .
\end{aligned} \end{equation}
The dispersion relations are then obtained by solving 
\begin{equation} \begin{aligned} \label{7.8}
	\omega \left( \omega + i \tau_{\text{gap}}^{-1} \right) - k^2 \approx 0 \Leftrightarrow 2 \omega \approx - i \tau_{\text{gap}}^{-1} \pm \sqrt{4 k^2 - \tau_{\text{gap}}^{-2}} \, .
\end{aligned} \end{equation}
We are interested in solutions for which $k / T \sim \varepsilon^{1 + \delta \kappa}$ and $\tau_{\text{gap}} T \sim \varepsilon^{-1}$ with $|\delta \kappa| \ll 1$ and $\varepsilon \ll |\delta \kappa|$.\footnote{Consequently, $\omega \sim \varepsilon^{1 + \delta \mathfrak{w}}$ for $|\delta \mathfrak{w}| \ll 1$.} When $\delta \kappa > 0$, we have one hydrodynamic and one quasihydrodynamic modes:
\begin{equation} \begin{aligned} \label{7.9}
	\omega = - i \tau_{\text{gap}} k^2 + O \left( \varepsilon^{1 + 4 \delta \kappa} T \right) \quad \quad \quad \text{and} \quad \quad \quad
	\omega = {- i \over \tau_{\text{gap}}} + i \tau_{\text{gap}} k^2 + O \left( \varepsilon^{1 + 4 \delta \kappa} T \right) .
\end{aligned} \end{equation}
The first dispersion relation is diffusive, while the second encodes purely damped relaxation with a diffusive $k^2$ correction that softens the damping rate.
On the other hand, when $\delta \kappa < 0$, we have two quasihydrodynamic modes:
\begin{equation} \begin{aligned} \label{7.10}
	\omega & = \pm k - {i \over 2 \tau_{\text{gap}}} + O \left( \varepsilon^{1 - \delta \kappa} T \right) .
\end{aligned} \end{equation}
These are \textit{relaxed sound} modes, valid for $k > \tau_{\text{gap}}^{-1}$. Regarding the overall structure of the low-energy dynamics prescribed by \eqref{eq:7.7}, we find that the regimes $k \ll \tau_{\text{gap}}^{-1}$ and $\tau_{\text{gap}}^{-1} \ll k \ll T$ encode the physics of diffusion and sound, respectively. Moreover, the dispersion relation becomes non-analytic in the intermediate, transient regime $k \sim \tau_{\text{gap}}^{-1}$, which is characterised by a \textit{pole collision}. 

\subsection{Constitutive relations and Josephson equations} \label{holog_interp}

\paragraph{}The theories defined by the massless path integrals \eqref{Za_p2} and \eqref{Zj_p2} both include the most relevant (non-topological) double-trace deformation allowed, parametrised by the coupling constants $\mathcal{M}$ and $\mathcal{M}^\prime$, respectively. 
In these cases, the source is dual to either
\begin{equation} \begin{aligned} \label{4.13_p2}
	a_{\mu_1 ... \mu_p} = \alpha_{\mu_1 ... \mu_p} + \mathcal{M} J_{\mu_1 ... \mu_p} \qquad \text{or} \qquad
	j_{\mu_1 ... \mu_p} = J_{\mu_1 ... \mu_p} - \mathcal{M}^\prime \partial^{\mu_0} \beta_{\mu_0 ... \mu_p} \, .
\end{aligned} \end{equation}
We extend convention \hyperlink{conv2}{1} to these fields, such that for example $a$ and $j$ tacitly refer to the $a_{A_1 ... A_p}$ and $j_{A_1 ... A_p}$ components.
When $\mathcal{M}$ and $\mathcal{M}^\prime$ vanish, \cref{7.4,7.7} can immediately be interpreted either as constitutive relations or Josephson equations of the undeformed boundary theory --- for a given set of sources, these determine expectation values in a derivative expansion. In general, however, these equations need to be written in terms of the \textit{final variables}.

\paragraph{}Starting with the theory defined by $\bar{Z}^{[{\bar{\lambda}} , \mathcal{M}]}$ corresponding to the electric quantisation of a massless $p$-form, \cref{7.4} is equivalent to
\begin{equation} \begin{aligned} \label{7.11a} 
	\mathcal{M} \Bigg( {1 \over \mathcal{M}} + r_h^{{\bar{\lambda}}-2} \partial_{t} + {r_h^{{\bar{\lambda}}-3} \over {\bar{\lambda}}-3} \partial_{x}^2 & + O (\omega , k^2) (\mathcal{M}^{-1} , \Box) \Bigg) J \\
	& = f_{t} r_h^{{\bar{\lambda}}-2} + \partial_{x} f_{x} {r_h^{{\bar{\lambda}}-3} \over {\bar{\lambda}}-3} + O (\omega , k^2) \partial^{\mu} f_{\mu} \, ,
\end{aligned} \end{equation}
where we introduced $f \vcentcolon = \mathrm{d} a / p!$ and used, in particular, that $\beta_\mu = f_\mu - \mathcal{M} \partial_\mu J$.\footnote{Our notation is such that $O (\omega , k^2) (\mathcal{M}^{-1} , \Box)$ stands for terms of order $O (\mathcal{M}^{-1} \omega , \mathcal{M}^{-1} k^2 , \omega^3 , \omega k^2 , k^4)$.} The holographic dictionary for such a theory dictates that in the large-$N$ limit $J^{\mu_1 ... \mu_p}$ is the expectation value of a conserved form-valued operator and $a_{\mu_1 ... \mu_p}$ is the conjugate source. Hence, for vanishing sources, the equation above is solved by plane-waves with non-hydrodynamic dispersion relations. In particular, it is of the form \eqref{quasihydro} when $|\mathcal{M}| \sim \tau_{\text{gap}}$ is parametrically large, in which case the dispersion relation is quasihydrodynamic.

\paragraph{}For the same theory, \cref{7.7} is equivalent to 
\begin{equation} \begin{aligned} \label{7.11b} 
	J_x r_h^{2-{\bar{\lambda}}} + \partial_{x} J_t {r_h^{1-{\bar{\lambda}}} \over 1-{\bar{\lambda}}} + \mathcal{M} \partial_{[t} J_{x]} + O (\omega , k^2) \left( 1 , \mathcal{M} \right) \partial_{[t} J_{x]} & = f_{t x} + O (\omega , k^2) f_{t x} \, ,
\end{aligned} \end{equation}
where $\beta_{\mu \nu} = f_{\mu \nu} - \mathcal{M} \partial_{[\mu} J_{\nu]}$ was used. The equation above is a constitutive relation for $J_x$ in terms of $J_t$ and background sources, that ceases to be local when $|\mathcal{M}|$ gets parametrically large. In that case, it can be written in the form \eqref{eq:7.7}.

\paragraph{}The two equations above follow from the left-most equation in \eqref{4.13_p2} (appropriate to electric quantisation). Using \eqref{1.38_p2} on these and comparing with the right-most equation in \eqref{4.13_p2} (appropriate to magnetic quantisation), we find that the Hodge-map
\begin{equation} \begin{aligned} \label{Hodge_renormalised}
	f^{({\bar{\lambda}})} & \to (-1)^{p+1} \ast j^{(4-{\bar{\lambda}})} \qquad \text{and} \qquad J^{({\bar{\lambda}})} \to \ast \beta^{(4-{\bar{\lambda}})} 
\end{aligned} \end{equation}
is valid for the \textit{final variables} of dual theories --- cf. \eqref{HodgeZ_massless_p2}. Hence, in the magnetic theory defined by $\bar{Z}^{[4-{\bar{\lambda}} , \mathcal{M}^\prime]}$, the equations analogous to \eqref{7.11a} and \eqref{7.11b} can be obtained from applying the map above (together with $\mathcal{M} \to \mathcal{M}^\prime$) and are therefore equivalent to them. In particular, one obtains two Josephson equations: one for the gauge-invariant part of $\alpha_{\mu = t , x}$, which is captured by $\beta_{t x}$, and another one for $\alpha$; from \eqref{7.11a} and \eqref{7.11b}, respectively. Note that, according to the dictionary for the current theory at large-$N$, $\alpha_{\mu_1 ... \mu_p}$ is the expectation value of a gauge non-invariant operator, conjugate to a conserved source $j^{\mu_1 ... \mu_p}$.

\paragraph{}We are not going to display the equation dual to \eqref{7.11a}, with the only point worth highlighting being that it also becomes a quasihydrodynamic equation of the form \eqref{quasihydro} when $|\mathcal{M}^\prime| \sim \tau_{\text{gap}}$ is parametrically large. The equation dual to \eqref{7.11b}, on the other hand, is worth presenting:\footnote{One should recall that, without resorting to Hodge duality, this would have been obtained by substituting $J = j + \mathcal{M}^\prime \partial^\mu f_\mu$ in \cref{7.4}.}
\begin{equation} \begin{aligned} \label{7.14b} 
	\beta_{t} r_h^{{\bar{\lambda}}-2} + \partial_{x} \beta_{x} {r_h^{{\bar{\lambda}}-3} \over {\bar{\lambda}}-3} - \mathcal{M}^\prime \partial^{\mu} \beta_\mu + O (\omega , k^2) \left( 1 , \mathcal{M}^\prime \right) \partial^{\mu} \beta_\mu & = j + O (\omega , k^2) j \, .
\end{aligned} \end{equation}
As was mentioned above, this is a Josephson equation for $\alpha$, which is gauge invariant due to the alignment of the wavevector. In particular, we can interpret it, supposing $\alpha$ is a charge density, as a conservation equation sourced by the term on the right-hand side:
\begin{equation} \begin{aligned} \label{9.23}
	\partial_t \alpha + \partial_x \left[ \left( {r_h^{-1} \over {\bar{\lambda}}-3} - {\mathcal{M}^\prime \over r_h^{{\bar{\lambda}}-2}} \right) \partial_x \alpha + \ldots \right] = - {\mathcal{M}^\prime \over r_h^{{\bar{\lambda}}-2}} \partial_t^2 \alpha + \ldots \, ,
\end{aligned} \end{equation} 
where we have ignored the sources and higher-order corrections. Naturally, by introducing the topological conservation equation (that follows from the Bianchi identity $\partial_{[\mu} \beta_{\nu]} = 0$), one can view \eqref{7.14b} as a constitutive relation such that the remarks below \eqref{7.11b} also apply. However, \cref{9.23} provides an alternative take on the origin of the quasihydrodynamic behaviour associated with $\alpha$ when $|\mathcal{M}^\prime|$ becomes large.

\paragraph{}Lastly, for the approximate conservation \cref{0.10} associated with a weakly broken higher-form symmetry to become a quasihydrodynamic equation of the form \eqref{quasihydro}, $\braket{\tilde{\mathcal{O}}^{i_2 \ldots i_p}}$ has to be given by a constitutive relation which is linear in $\braket{\mathcal{O}^{t i_2 \ldots i_p}}$. Accordingly, \eqref{6.57} is a quasihydrodynamic equation for $K_t^{\mp}$ upon taking \eqref{6.45} into account.

\subsubsection{Higher-form electromagnetism at the boundary} \label{electromag}

\paragraph{}\Cref{7.11b,7.14b} entail the existence of a diffusion-to-sound crossover, as discussed in \cref{Quasihydro_sec}, in the ${\bar{\lambda}}_{\text{eff}} = 1$ sector of the electric theory and in the ${\bar{\lambda}}_{\text{eff}} = 3$ sector of the magnetic theory. This naturally concerns regimes of low frequency and wavenumber with respect to the temperature. \\ \hspace*{1.75ex}
Let us focus on the relaxed sound modes that populate the aforementioned sectors for 
\begin{equation} \begin{aligned}
	\mathcal{M} k \gg r_h^{2-{\bar{\lambda}}} \qquad \quad \text{and} \qquad \quad \mathcal{M}^\prime k \gg r_h^{{\bar{\lambda}}-2} \, .
\end{aligned} \end{equation}
We are going to show that they are governed by the equations of higher-form electromagnetism at leading order in 
\begin{equation} \begin{aligned} \label{9.25}
	r_h^{1-{\bar{\lambda}}} / \mathcal{M} \ll 1 \qquad \quad \text{and} \qquad \quad r_h^{{\bar{\lambda}}-3} / \mathcal{M}^\prime \ll 1 \, .
\end{aligned} \end{equation} 
Note that \cref{7.11b,7.14b} can be written respectively as
\begin{subequations} \label{9.26} \begin{align} 
	\label{9.26a} \partial_{[t} J_{x]} \left( 1 + \ldots \right) & = \mathcal{M}^{-1} f_{t x} \\
	\label{9.26b} \partial^\mu \beta_{\mu} \left( 1 + \ldots \right) & = - (\mathcal{M}^\prime)^{-1} j \, ,
\end{align} \end{subequations}
where the ellipses stand for corrections associated with \eqref{9.25} and higher-derivative corrections have been completely omitted. Viewing $\beta_\mu$ as the field strength associated with the ${A_1 ... A_{p}}$ components of a $p$-form electromagnetic potential, \eqref{9.26b} reproduces the corresponding components of the Maxwell equation of motion in flat Minkowski space. These equations are sourced by the external $p$-form electric current $j$ and $(\mathcal{M}^\prime)^{-1}$ is an effective gauge coupling. Note that $\alpha_{\mu = t , x}$, unlike $\alpha$, is not governed by flat-space Maxwell equations. 

\paragraph{}Similarly, for the boundary theory in electric quantisation, the conservation equation $\partial_\mu J^\mu = 0$ and \eqref{9.26a} reproduce the flat-space Maxwell equations:\footnote{This time, in first-order formalism.} viewing $J_{\mu_1 ... \mu_p} - \mathcal{M}^{-1} a_{\mu_1 ... \mu_p}$ as the field strength, these are the ${A_2 ... A_{p}}$ and ${t x A_2 ... A_{p}}$ components of the equation of motion and Bianchi identity, respectively. The external electric current in this case is $\partial_{\mu_1} a^{\mu_1 ... \mu_p}$ and the gauge coupling is $\mathcal{M}^{-1}$.\footnote{Alternatively, one can 1) Hodge dualise and introduce a dual electromagnetic potential to interpret it in the same manner as \cref{9.26b}; or 2) consider a free Maxwell equation of motion and a ``Bianchi identity" that is sourced by $f_{t x}$ as an external current.} Note that the Maxwell equations govern $J_{\mu = t , x}$ but not $J$.

\section{Low-energy structure of thermal correlators} \label{correlator_results}

\paragraph{}In what follows, we use the results from \cref{hydrolim} to derive 2-point connected thermal correlators in all of the boundary theories we are considering --- this includes the holographic duals to both quantisations of the massless and massive bulk fields. Such correlators are schematically given by $\braket{\mathcal{O} \mathcal{O}}_{\text{con}} \equiv \braket{\mathcal{O} \mathcal{O}} - \braket{\mathcal{O}} \braket{\mathcal{O}}$ and correspond to the second derivative of the logarithm of the respective path integral, cf. \cref{Za_p2,Zpm_p2,Zj_p2,Zmp}.
Let us start with electric quantisation in the massless case, where we write 
\begin{equation} \begin{aligned} \label{firstderiv}
	{\delta \ln \bar{Z}^{[{\bar{\lambda}} , \mathcal{M}]} \over \delta \psi_{\mu_1 ... \mu_p}} = {\int_{a [A] \vert_{\partial \mathbb{B}} = \psi} \mathfrak{D} A e^{i \bar{S}_{\text{final}}} J^{\mu_1 ... \mu_p} [A] \big\vert_{\partial \mathbb{B}} \over \bar{Z}^{[{\bar{\lambda}} , \mathcal{M}]}} \, ,
\end{aligned} \end{equation}
after introducing the functionals $a [A]$ and $J [A]$ that approach the \textit{final variables} at the boundary when $A$ is on-shell.\footnote{Recall footnote \ref{integ_domain} in \cref{p2setup}. Also, see eqs. (2.63) in \cite{Pinheiro:2026ihi}.} We consider transverse ($A_1 ... A_p$) and longitudinal components ($t A_2 ... A_p$ and $x A_2 ... A_p$), for concreteness. \\ \hspace*{1.75ex}
The right-hand side of the equation above is the normalised insertion of $J [A] \big\vert_{\partial \mathbb{B}}$ in the path integral, which in the large-$N$ limit becomes $J [A] \big\vert_{\partial \mathbb{B}}$ evaluated in a classical configuration obeying boundary conditions $a [A] \big\vert_{\partial \mathbb{B}} = \psi$ and regularity conditions in the interior of $\mathbb{B}$. We consider regularity in coordinates adapted to radial ingoing null geodesics, which results in solutions approaching the horizon in coordinates $(\rho, x^\mu)$ (discussed near the end of \cref{nearhorizon}) as an ingoing wave. Hence, in the large-$N$ limit, the right-hand side of \eqref{firstderiv} is a function $J^{\mu_1 ... \mu_p}$ of $\psi$ found by solving equations \eqref{7.4} or \eqref{7.7}, upon solving $a = \psi$ for $\alpha$. To be specific, the function $J$ obtained is actually the Fourier transform of the path integral insertion in the hydrodynamic (low-frequency, low-wavenumber) regime. Differentiating $J$ at $\psi = \hat{\psi}$, one obtains the large-$N$ limit of the connected correlator in Fourier space:
\begin{equation} \begin{aligned} \label{4.15a} 
	 \braket{\bar{\mathcal{O}}_j^{\mu_1 ... \mu_p} \bar{\mathcal{O}}_j^{\nu_1 ... \nu_p}}_{\mathrm{R}} & \xrightarrow[N \to \infty]{} - i {\partial J^{\mu_1 ... \mu_p} \over \partial a_{\nu_1 ... \nu_p}} \bigg\vert_{a = \hat{\psi}} ,
\end{aligned} \end{equation}
where it is assumed that bulk path integrals are normalised such that, in the present case, $\bar{Z}^{[{\bar{\lambda}} , \mathcal{M}]} \big\vert_{\psi = \hat{\psi}} = 1$. (This can always be achieved by an appropriate rescaling of the integration measure). Due to our choice of regularity conditions, these are the retarded thermal correlators \cite{Son:2002sd,Herzog:2002pc} --- hence the label $\mathrm{R}$. A similar result holds for the remaining boundary theories. In particular, for the magnetic quantisation of a massless bulk field, the second derivative of $\ln \bar{Z}^{[{\bar{\lambda}} , \mathcal{M}^\prime]}$ at $\psi = \hat{\psi}$ leads to 
\begin{equation} \begin{aligned} \label{4.15b} 
	 \braket{\bar{\mathcal{O}}^a_{\mu_1 ... \mu_p} \bar{\mathcal{O}}^a_{\nu_1 ... \nu_p}}_{\mathrm{R}} & \xrightarrow[N \to \infty]{} - i {\partial \alpha_{\mu_1 ... \mu_p} \over \partial j^{\nu_1 ... \nu_p}} \bigg\vert_{j = \hat{\psi}} .
\end{aligned} \end{equation}
Note that, in keeping with the use of bars throughout the paper to distinguish the massless case, we have introduced $\bar{\mathcal{O}}_j$ and $\bar{\mathcal{O}}^a$ above. \\ \hspace*{1.75ex}
In the massive case, the first derivative of $Z^{[\lambda , \mathcal{M}]}$ in the large-$N$ limit is the Fourier transform of a function $K_{\mp}$ of $\psi$ found by solving \cref{9.27,6.45}, upon solving $\mathcal{K}_{\pm} = \psi$ for $K_{\pm}$. Going from electric to magnetic quantisation, the same statement can be made if we exchange labels $\mp \leftrightarrow \pm$. Hence, the connected thermal correlators in Fourier space are
\begin{subequations} \label{4.42} \begin{align}
	\label{4.42a}  \braket{\mathcal{O}_j^{\mu_1 ... \mu_p} \mathcal{O}_j^{\nu_1 ... \nu_p}}_{\mathrm{R}} & \xrightarrow[N \to \infty]{} - i {\partial K_{\mp}^{\mu_1 ... \mu_p} \over \partial \mathcal{K}^{\pm}_{\nu_1 ... \nu_p}} \bigg\vert_{\mathcal{K}^{\pm} = \hat{\psi}} \\
	\label{4.42b}  \braket{\mathcal{O}_a^{\mu_1 ... \mu_p} \mathcal{O}_a^{\nu_1 ... \nu_p}}_{\mathrm{R}} & \xrightarrow[N \to \infty]{} - i {\partial K_{\pm}^{\mu_1 ... \mu_p} \over \partial \mathcal{K}^{\mp}_{\nu_1 ... \nu_p}} \bigg\vert_{\mathcal{K}^{\mp} = \hat{\psi}} .
\end{align} \end{subequations}

\paragraph{}Convention \hyperlink{conv2}{1} was clearly not in use above, but it is reinstated for the rest of the paper. Also, we will be using the dimensionless quantities $\hat{\omega}$ and $\hat{k}$ introduced in the previous section. These are normalised by $r_h$ which, for black branes with hair, generally depends on multiple scales, including temperature and chemical potentials.
In light of the discussion in \cref{Quasihydro_sec}, this suggests that the microscopic scale signalling the breakdown of the (quasi)hydrodynamic analysis is set by the entropy density (which is equivalent to the horizon area), rather than by the temperature itself. However, one must bear in mind that, for many families of solutions in different models, $r_h$ approaches $T$ when the chemical potentials (or, rather, their thermodynamic conjugates) are low, which is precisely the regime in which the probe limit we consider is a reliable approximation.

\subsection{Two-point retarded correlators: the massless case} \label{masslesscorrelators}

\paragraph{}According to the discussion above, the results from \cref{hydrolim} allow us to compute all the non-trivial retarded correlators, which group into pairs according to Hodge duality:
\begin{equation} \begin{aligned} \label{9.33}
	\Big\lbrace  \braket{\bar{\mathcal{O}}_j^{A_1 ... A_p} \bar{\mathcal{O}}_j^{A_1 ... A_p}}_{\mathrm{R}} , & \braket{\bar{\mathcal{O}}^a_{\mu A_2 ... A_p} \bar{\mathcal{O}}^a_{\nu A_2 ... A_p}}_{\mathrm{R}} \Big\rbrace \\ 
	& \qquad \qquad \qquad \text{and} \; \; \Big\lbrace \braket{\bar{\mathcal{O}}_j^{\mu A_2 ... A_p} \bar{\mathcal{O}}_j^{\nu A_2 ... A_p}}_{\mathrm{R}} , \braket{\bar{\mathcal{O}}^a_{A_1 ... A_p} \bar{\mathcal{O}}^a_{A_1 ... A_p}}_{\mathrm{R}} \Big\rbrace .
\end{aligned} \end{equation}
where $\mu , \nu \in \{ t , x \}$. Let us introduce dimensionless coupling constants, $\hat{\mathcal{M}} = r_h^{{\bar{\lambda}}-1} \mathcal{M}$ and $\hat{\mathcal{M}}^\prime = r_h^{3-{\bar{\lambda}}} \mathcal{M}^\prime$, for the double-trace deformations. 

\subsubsection{Hydrodynamic regime}

\paragraph{}The correlators of the second pair in \eqref{9.33} have diffusive poles in the large-$N$ limit when $|\hat{\mathcal{M}}^{(\prime)} \hat{k}| \ll 1$.\footnote{A more precise bound on $\hat{\mathcal{M}}^{(\prime)} \hat{k}$ is given below.} However, as we shall see below, the low-energy spectrum associated with these correlators also contain quasihydrodynamic modes when $|\hat{\mathcal{M}}^{(\prime)}| \gg 1$. For now, we are interested in the hydrodynamic regime and therefore restrict to $|\hat{\mathcal{M}}^{(\prime)}| \lesssim 1$. The correlators in this regime are given, according to \eqref{4.15a} and \eqref{4.15b}, by
\begin{subequations} \label{6.32} \begin{align} 
	\label{6.32a} {\delta J^\mu \over \delta a_\nu} & = {- i r_h^{{\bar{\lambda}}-1} + O (\varepsilon^2) \over \hat{\omega} + i \left( {1 \over {\bar{\lambda}}-1} + \hat{\mathcal{M}} \right) \hat{k}^2 + O (\varepsilon^4 \hat{\mathcal{M}} , \varepsilon^4)} 
	\begin{blockarray}{cccc}
		& \text{\tiny $\mu = t$} & \text{\tiny $\mu = x$} & \\
		\begin{block}{c(cc)c}
			& \hat{k}^2 & \hat{\omega} \hat{k} & \text{\tiny $\nu = t$} \\
			& \hat{\omega} \hat{k} & \hat{\omega}^2 & \text{\tiny $\nu = x$} \\
		\end{block}
	\end{blockarray} \\
	\label{6.32b} {\delta \alpha \over \delta j} & = {i r_h^{1-{\bar{\lambda}}} + O (\varepsilon^2) \over \hat{\omega} + i \left( {1 \over 3-{\bar{\lambda}}} + \hat{\mathcal{M}}^\prime \right) \hat{k}^2 + O (\varepsilon^4 \hat{\mathcal{M}}^\prime , \varepsilon^4)} \, ,
\end{align} \end{subequations}
where we have assumed that $\hat{k} \sim \varepsilon$ and $\hat{\omega} \sim \varepsilon^2$ in order to simplify our presentation of the subleading terms' order. The components of the matrix on the right-hand side of \cref{6.32a} match $(- r_h^{-2} k^\rho k_\rho) \Pi^{\mu \nu}$, where $\Pi^{\mu \nu} = \eta^{\mu \nu} - k^\mu k^\nu / k^\rho k_\rho$ is the transverse projector, as to be expected from the Ward identities.

\paragraph{}The low-energy poles in \eqref{6.32a} and \eqref{6.32b} correspond to dispersion relations of diffusive modes, for which the diffusion constant is given respectively by
\begin{equation} \begin{aligned} \label{9.35}
	D = r_h^{{\bar{\lambda}}-2} \left( \mathcal{M} - {r_h^{1-{\bar{\lambda}}} \over 1-{\bar{\lambda}}} \right) \qquad \quad \text{and} \qquad \quad D^\prime = r_h^{2-{\bar{\lambda}}} \left( \mathcal{M}^\prime - {r_h^{{\bar{\lambda}}-3} \over {\bar{\lambda}}-3} \right) ,
\end{aligned} \end{equation}
where notation \eqref{6.22} applies for ${\bar{\lambda}}=1$ and ${\bar{\lambda}} = 3$. Positivity of the diffusion constant ensures linear stability in the hydrodynamic regime. From our results, we see that this cannot be achieved without a deformation when ${\bar{\lambda}} < 1$ for electric theories and ${\bar{\lambda}} > 3$ for magnetic ones.\footnote{For $r_h > 1$, this statement can be extended to ${\bar{\lambda}} \leq 1$ and ${\bar{\lambda}} \geq 3$.} These are precisely the ranges where the double-trace deformations considered here are relevant.

\paragraph{}A convenient way to see that the correlators under consideration are related by Hodge duality is to first dress them with appropriate powers of $\hat{\omega}$ and $\hat{k}$, and then form linear combinations of different components such that $a_{\mu_1 ... \mu_p}$ and $\alpha_{\mu_1 ... \mu_p}$ in \eqref{6.32} appear only through the respective field strengths. At this stage, one simply applies the Hodge map in the form given in \eqref{Hodge_renormalised}.
At the level of poles, however, the duality is immediate. In particular, the diffusivity $D^\prime$ can be obtained from $D$ via the replacement $\{ {\bar{\lambda}} \to 4 - {\bar{\lambda}} , \hat{\mathcal{M}} \to \hat{\mathcal{M}}^\prime \}$, in agreement with \eqref{HodgeZ_massless_p2}.

\subsubsection{Quasihydrodynamic regime}

\paragraph{}We now turn to the large-$N$ limit of the first pair of correlators in \eqref{9.33}, given by
\begin{subequations} \label{9.37} \begin{align} 
	\label{630} {\delta J \over \delta a} & = {r_h^{{\bar{\lambda}}-1} \over \hat{\mathcal{M}}} {\hat{\omega} + i {\hat{k}^2 \over 3-{\bar{\lambda}}} + O (\varepsilon^4) \over \hat{\omega} + {i \over \hat{\mathcal{M}}} + i {\hat{k}^2 \over 3-{\bar{\lambda}}} + O (\varepsilon^2 / \hat{\mathcal{M}} , \varepsilon^4)} \\
	\label{631} 
	\begin{pmatrix}
		\hat{k} {\delta \alpha_{t} \over \delta j^{t}} & \hat{\omega} \\
		\hat{\omega} {\delta \alpha_{t} \over \delta j^{x}} & \hat{\omega} 
	\end{pmatrix}
	\begin{pmatrix}
		\hat{k} & \hat{k} \\
		\hat{k} {\delta \alpha_{x} \over \delta j^{t}} & \hat{\omega} {\delta \alpha_{x} \over \delta j^{x}}
	\end{pmatrix} 
	& = {r_h^{1-{\bar{\lambda}}} \over \hat{\mathcal{M}}^\prime} {\hat{\omega} + i {\hat{k}^2 \over {\bar{\lambda}}-1} + O (\varepsilon^4) \over \hat{\omega} + {i \over \hat{\mathcal{M}}^\prime} + i {\hat{k}^2 \over {\bar{\lambda}}-1} + O (\varepsilon^2 / \hat{\mathcal{M}}^\prime , \varepsilon^4)}
	\begin{pmatrix}
		1 & 1 \\
		1 & 1 
	\end{pmatrix} \, .
\end{align} \end{subequations}
Note that the left-hand side of the bottom equation is a matrix of gauge-invariant combinations. Even though these expressions only have $\hat{\omega} \sim \hat{k}^2$ poles when $|\hat{\mathcal{M}}^{(\prime)} \hat{k}^2| \gtrsim 1$, we have presented the order of subleading terms under the assumption that $\hat{k} \sim \varepsilon$ and $\hat{\omega} \sim \varepsilon^2$, for simplicity.

\paragraph{}For positive $\hat{\mathcal{M}}^{(\prime)} \gg 1$, the poles in \eqref{9.37} exhibit \textit{relaxation}, by which we mean that the leading contributions to the dispersion relations $\omega (k)$ correspond to (pure) damping at a slow rate plus diffusion. Interestingly, the diffusivities in this case coincide with ${D}^{(\prime)}$ if we set $\mathcal{M}^{(\prime)} = 0$ by hand in \eqref{9.35}. The damping term, which places a small lower bound on $|\Im \omega (k)|$ as we saw in \cref{Quasihydro_sec}, is associated with $\tau_{\text{gap}} = {\hat{\mathcal{M}}^{(\prime)} / r_h}$. Note that the current modes do not become hydrodynamic in the $\hat{\mathcal{M}}^{(\prime)} \to \infty$ limit, since in this case their dispersion relation ceases to be a low-energy pole of the respective correlators (as the numerator and denominator cancel each other out).\footnote{The matching between the numerator's and denominator's $O (\varepsilon^4)$ terms ensures true cancellation. One can check this, for example in electric quantisation, from the way terms proportional to $\mathcal{M}$ enter \cref{7.11a} via $\beta \equiv f - \mathcal{M} \mathrm{d} J / p!$.}

\paragraph{}In \eqref{6.32}, we examined the remaining pair of large-$N$ correlators in the hydrodynamic regime $|\hat{\mathcal{M}}^{(\prime)}| \lesssim 1$. By contrast, in the quasihydrodynamic regime $|\hat{\mathcal{M}}^{(\prime)}| \sim \varepsilon^{-1} \gg 1$, they are given by
\begin{subequations} \begin{align} 
	\label{6.35a} {\delta J^\mu \over \delta a_\nu} & = {r_h^{{\bar{\lambda}}-1} / \hat{\mathcal{M}} + O (\varepsilon \omega , \varepsilon k^2) \over \hat{\omega} \left( \hat{\omega} + {i \over \hat{\mathcal{M}}} \right) - \hat{k}^2 + O (\hat{\omega}^3, \hat{\omega} \hat{k}^2 , \hat{k}^4 , \varepsilon \hat{k}^2)}
	\begin{blockarray}{cccc}
		& \text{\tiny $\mu = t$} & \text{\tiny $\mu = x$} & \\
		\begin{block}{c(cc)c}
			& \hat{k}^2 & \hat{\omega} \hat{k} & \text{\tiny $\nu = t$} \\
			& \hat{\omega} \hat{k} & \hat{\omega}^2 & \text{\tiny $\nu = x$} \\
		\end{block}
	\end{blockarray} \\
	\label{6.35b} {\delta \alpha \over \delta j} & = {- r_h^{1-{\bar{\lambda}}} / \hat{\mathcal{M}}^\prime + O (\varepsilon \omega , \varepsilon k^2) \over \hat{\omega} \left( \hat{\omega} + {i \over \hat{\mathcal{M}}^\prime} \right) - \hat{k}^2 + O (\hat{\omega}^3, \hat{\omega} \hat{k}^2 , \hat{k}^4 , \varepsilon \hat{k}^2)} \, .
\end{align} \end{subequations}
These expressions have poles given by \cref{7.8} with $\tau_{\text{gap}} = {\hat{\mathcal{M}}^{(\prime)} / r_h}$. Hence, assuming that $\hat{k} \sim \varepsilon^{1 + \delta \kappa}$ where $\varepsilon \ll \delta \kappa \ll 1$, we obtain a pair of dispersion relations \eqref{7.9} encoding relaxation and diffusion. While the former is given by
\begin{equation} \begin{aligned} 
	\hat{\omega} & = {- i \over \hat{\mathcal{M}}^{(\prime)}} + i \hat{\mathcal{M}}^{(\prime)} \hat{k}^2 + O \left( \varepsilon^{1 + 4\delta \kappa} \right) ,
\end{aligned} \end{equation}
the latter corresponds to the previously discussed hydrodynamic mode with diffusivity \eqref{9.35}. If, on the other hand, $\varepsilon \ll - \delta \kappa \ll 1$, then \eqref{7.10} implies a pair of relaxed sound modes with dispersion relations
\begin{equation} \begin{aligned} \label{6.37b} 
	\hat{\omega} & = \pm \hat{k} - {i \over 2 \hat{\mathcal{M}}^{(\prime)}} + O \left( \varepsilon^{1 - \delta \kappa} \right) .
\end{aligned} \end{equation}
Note that the $O \left( \varepsilon^{1 - \delta \kappa} \right)$ term corrects the speed of sound. In agreement with the electromagnetic behaviour discussed in \cref{electromag}, in the large-$\hat{\mathcal{M}}^{(\prime)}$ limit, the relaxation time of the sound modes becomes infinite and they propagate at the speed of light: $\hat{\omega} = \pm \hat{k}$. In total, one sees that $\binom{d-2}{p-1}$ and $\binom{d-2}{p}$ photon polarisations emerge in the electric and magnetic theories, respectively.\footnote{This generalises the photon found in the context of $d=4$ magnetohydrodynamics by \cite{Grozdanov:2018fic}, with a $p{=}2$-form potential in the bulk.} A crucial point is that these modes fully account for the low-energy spectrum in this regime, since the correlators associated with \eqref{9.37} do not contribute in the large-$\hat{\mathcal{M}}^{(\prime)}$ limit. Hence, even though equations \eqref{9.26} do not in general comprise all components of the flat-space Maxwell equations,\footnote{The exceptions are the electric theory where $p= d - 1$ and the magnetic theory where $p= 0$. In these cases, the first pair of correlators in \eqref{9.33} does not exist.} our holographic theories still yield the same propagating degrees of freedom.

\paragraph{}In general, the spectra (for a given ${\bar{\lambda}}$) in different quantisations are qualitatively identical provided that $\hat{\mathcal{M}}$ and $\hat{\mathcal{M}}^\prime$ are of the same order $O (\varepsilon^{\mathtt{n}})$, $\mathtt{n} \in \mathbb{Z}$.\footnote{Assuming that $\hat{\mathcal{M}} > {1 \over 1-{\bar{\lambda}}}$ and $\hat{\mathcal{M}}^\prime > {1 \over {\bar{\lambda}}-3}$, as a condition for linear stability.} The obvious exceptions are the lowest and highest-rank cases of a scalar and a $(d{-}1)$-form potential for which half of the correlators in \eqref{9.33} do not exist. 

\subsection{Two-point retarded correlators: the massive case} \label{massivecorrelators}

\paragraph{}In \cref{hydrolim}, we also derived equations \eqref{9.27} which relate $K^{\pm}$ to $K^{\mp}$ and $K^{\mp}_{t x}$ to $K^{\pm}_{t x}$. These stem from the $\lambda_{\text{eff}} = 5$ and $\lambda_{\text{eff}} = 1$ systems, which are dual under massive Hodge duality --- cf. \eqref{Hodgemap2_p2}. The relation between $K^{\pm}_{\mu = t , x}$ and $K^{\mp}_{\mu = t , x}$ was also determined and is given, in the scaling limit where $\hat{\omega} \sim \hat{k} \sim m$, by \cref{7.47}. This one arises from the self-dual $\lambda_{\text{eff}} = 3$ system. While we restrict to $\hat{k} \sim m$ such that we are probing the system at the defect characteristic scale, below we will also consider the self-dual sector when $\hat{\omega} \sim \hat{k}^2 \sim m^2$. In this case, we use \cref{7.46} from the appendices.

\subsubsection{The dual sectors}

\paragraph{}We begin by using equations \eqref{9.27}, written in terms of the \textit{final variables} introduced in \eqref{4.28_p2}, to compute the following retarded correlators,
\begin{equation} \begin{aligned} \label{9.41}
	\Big\lbrace \braket{\mathcal{O}_j^{t x A_3 ... A_p} \mathcal{O}_j^{t x A_3 ... A_p}}_{\mathrm{R}} , & \braket{\mathcal{O}_a^{A_1 ... A_p} \mathcal{O}_a^{A_1 ... A_p}}_{\mathrm{R}} \Big\rbrace \\ 
	& \qquad \qquad \text{and} \; \; \Big\lbrace \braket{\mathcal{O}_j^{A_1 ... A_p} \mathcal{O}_j^{A_1 ... A_p}}_{\mathrm{R}} , \braket{\mathcal{O}_a^{t x A_3 ... A_p} \mathcal{O}_a^{t x A_3 ... A_p}}_{\mathrm{R}} \Big\rbrace ,
\end{aligned} \end{equation}
which were grouped into pairs according to Hodge duality. Once again, we use dimensionless coupling constants, $\hat{\mathcal{M}} = r_h^{\lambda-3} \mathcal{M}$ and $\hat{\mathcal{M}}^\ast = r_h^{3-\lambda} \mathcal{M}^\ast$, for the double-trace deformations.

\paragraph{}The correlators of the first pair in \eqref{9.41} exhibit quasihydrodynamic relaxation modes when $|\hat{\mathcal{M}}^\ast| \ll 1$ and $|\hat{\mathcal{M}} m^2| \ll 1$. They are given, according to \eqref{4.42}, by
\begin{subequations} \label{6.60} \begin{align} 
	\label{6.60b} {\delta K_\mp^{t x} \over \delta \mathcal{K}^\pm_{t x}} & = {- i r_h^{\lambda-3} {m^2 \over 3-\lambda} + O (\varepsilon^4) \over \hat{\omega} + i {\hat{k}^2 \over \lambda-1} + i {m^2 \over \lambda-3} \left( 1 + \hat{\mathcal{M}} \right) + O (\varepsilon^4 \mathcal{M} , \varepsilon^4)} \\
	\label{6.60a} {\delta K_{\pm} \over \delta \mathcal{K}^\mp} & = {- i r_h^{3-\lambda} (\lambda - 3) + O ( \varepsilon^2) \over \hat{\omega} + i {\hat{k}^2 \over 5-\lambda} + i {m^2 \over 3-\lambda} \left( 1 - {(\lambda - 3)^2 \over m^2} \hat{\mathcal{M}}^\ast \right) + O ( \varepsilon^2 \hat{\mathcal{M}}^\ast , \varepsilon^4)} \, .
\end{align} \end{subequations}
Here, we are simplifying the order of subleading terms by assuming that $\hat{k} \sim \varepsilon \sim m$ and $\hat{\omega} \sim \varepsilon^2$. Similarly, the correlators of the second pair yield relaxation modes when $|\hat{\mathcal{M}}^\ast / m^2| \gg 1$ and $|\hat{\mathcal{M}}| \gg 1$ and they are given by
\begin{subequations} \label{6.61} \begin{align}  
	\label{6.61b} {\delta K_{\mp} \over \delta \mathcal{K}^\pm} & = {- r_h^{\lambda-3} \over \hat{\mathcal{M}}} {\hat{\omega} + i {\hat{k}^2 \over 5-\lambda} + i {m^2 \over 3-\lambda} + O ( \varepsilon^4) \over \hat{\omega} + i {\hat{k}^2 \over 5-\lambda} + i {m^2 \over 3-\lambda} \left( 1 - {(\lambda - 3)^2 \over m^2 \hat{\mathcal{M}}} \right) + O (\varepsilon^2 / \hat{\mathcal{M}} , \varepsilon^4)} \\
	\label{6.61a} {\delta K_{\pm}^{t x} \over \delta \mathcal{K}^\mp_{t x}} & = {r_h^{3-\lambda} \over \hat{\mathcal{M}}^\ast} {\hat{\omega} + i {\hat{k}^2 \over \lambda-1} + i {m^2 \over \lambda-3} + O (\varepsilon^4) \over \hat{\omega} + i {\hat{k}^2 \over \lambda-1} + i {m^2 \over \lambda-3} \left( 1 + {1 \over \hat{\mathcal{M}}^\ast} \right) + O (\varepsilon^4 / \hat{\mathcal{M}}^\ast , \varepsilon^4)} \, .
\end{align} \end{subequations}
These share with \eqref{9.37} the fact that they do not have poles in the strong deformation limit.

\paragraph{}Recall that we are considering Robin boundary conditions $K^{\pm} - \mathcal{M} K^{\mp} = \psi$ for the electric theories and $K^{\mp} - \mathcal{M}^\ast K^{\pm} = \psi'$ for the magnetic ones.\footnote{The sources $\psi$ and $\psi'$ are arbitrary.} As a result, when $\mathcal{M} \mathcal{M}^\ast = 1$ and $\psi = - \mathcal{M} \psi'$, these theories are related by a strong/weak-coupling duality. Hence, on one hand we have massive Hodge duality, which implies equivalence of correlation functions for different quantisations and different $\lambda$. Specifically for the correlators at hand, it explains why \eqref{6.60a} and \eqref{6.61a} can be obtained by substituting 
\begin{equation} \begin{aligned} \label{9.44}
	\lambda \to 6 - \lambda \qquad \text{and} \qquad \hat{\mathcal{M}} \to - {(\lambda - 3)^2 \over m^2} \hat{\mathcal{M}}^\ast 
\end{aligned} \end{equation}
in \eqref{6.60b} and \eqref{6.61b}, respectively, and scaling these by ${(\lambda-3)^2 \over m^2}$.\footnote{This agrees with the small-$m^2$ limit of $- {\varDelta_\mp \over \varDelta_\pm}$ and therefore is consistent with equations \eqref{5.18_p2}.}
On the other hand, we have strong/weak duality ensuring equivalence, up to contact terms, of correlation functions for different quantisations but the same $\lambda$. Because contact terms do not affect the poles, one can check that, if we substitute $\hat{\mathcal{M}}^\ast \to 1 / \hat{\mathcal{M}}$ in equations \eqref{6.60a} and \eqref{6.61a}, we obtain modes with the same dispersion relations as
\eqref{6.61b} and \eqref{6.60b}, respectively.

\paragraph{}The dualities under consideration relate the correlators \eqref{9.41} according to the following diagram
\begin{center}
	\begin{tikzpicture}
		\node (E) at (0,0) {$\braket{\mathcal{O}_a^{A_1 ... A_p} \mathcal{O}_a^{A_1 ... A_p}}_{\mathrm{R}}$};
		\node
		(F) at (5,0) {$\braket{\mathcal{O}_j^{t x A'_3 ... A'_{p'}} \mathcal{O}_j^{t x A'_3 ... A'_{p'}}}_{\mathrm{R}}$};
		\node
		(N) at (5,-1.8) {$\braket{\mathcal{O}_a^{t x A'_3 ... A'_{p'}} \mathcal{O}_a^{t x A'_3 ... A'_{p'}}}_{\mathrm{R}}$};
		\node[below=of E] (M) {$\braket{\mathcal{O}_j^{A_1 ... A_p} \mathcal{O}_j^{A_1 ... A_p}}_{\mathrm{R}}$};
		\draw[<->] (E)--(F) node [midway,above] {};
		\draw[<->] (F)--(N) node [midway,right] {};
		\draw[<->] (M)--(N) node [midway,below] {};
		\draw[<->] (E)--(M) node [midway,left] {};
	\end{tikzpicture} 
\end{center}
where vertical arrows represent strong/weak duality and horizontal arrows represent Hodge duality, such that $p' = d - p$. As an example, note how the diagram above is realised by the pure damping term in the relevant dispersion relations:
\begin{center}
	\begin{tikzpicture}
		\node (E) at (0,0) {${m^2 \over 3-\lambda} \left( 1 - {(\lambda - 3)^2 \over m^2} \hat{\mathcal{M}}^\ast \right)$};
		\node[right=of E] (F) {${m^2 \over \lambda-3} \left( 1 + \hat{\mathcal{M}} \right)$};
		\node[below=of F] (N) {${m^2 \over \lambda-3} \left( 1 + {1 \over \hat{\mathcal{M}}^\ast} \right)$};
		\node[below=of E] (M) {${m^2 \over 3-\lambda} \left( 1 - {(\lambda - 3)^2 \over m^2 \hat{\mathcal{M}}} \right)$};
		\draw[<->] (E)--(F) node [midway,above] {};
		\draw[<->] (F)--(N) node [midway,right] {};
		\draw[<->] (M)--(N) node [midway,below] {};
		\draw[<->] (E)--(M) node [midway,left] {};
	\end{tikzpicture}
\end{center}
The composition of the two dualities maps between identical quantisations and, in particular, relates modes in the $\lambda_{\text{eff}} = 5$ and $\lambda_{\text{eff}} = 1$ sectors.

\subsubsection{The self-dual sector}

\paragraph{}The remaining non-trivial retarded correlators are
\begin{equation} \begin{aligned} \label{9.45}
	\braket{\mathcal{O}_j^{\mu A_2 ... A_p} \mathcal{O}_j^{\nu A_2 ... A_p}}_{\mathrm{R}} \qquad \text{and} \qquad \braket{\mathcal{O}_a^{\mu A_2 ... A_p} \mathcal{O}_a^{\nu A_2 ... A_p}}_{\mathrm{R}} \, , \qquad \mu , \nu \in \{ t , x \} \, ,
\end{aligned} \end{equation}
which are related by both massive Hodge duality and strong/weak-coupling duality. This time, for the sake of simplicity, we present only the results for the electric theory, which concern the correlator on the left-hand side. We also restrict to the case where $\hat{k} \sim \varepsilon \sim m$ and $m^2 > 0$.

\paragraph{}Using \eqref{7.47} in \cref{4.42}, the electric correlator in \eqref{9.45} is given by
\begin{equation} \begin{aligned} \label{6.71}
	{\delta K_\mp^\mu \over \delta \mathcal{K}^\pm_\nu} & = {- r_h^{\lambda-3} / \hat{\mathcal{M}} + O (\varepsilon^2) \over \hat{\omega} \left( \hat{\omega} + i \mathcal{C} \right) - \hat{k}^2 - m^2 + O (\varepsilon^3)} 
	\begin{blockarray}{cccc}
		& \text{\tiny $\mu = t$} & \text{\tiny $\mu = x$} & \\
		\begin{block}{c(cc)c}
			& \hat{k}^2 + m^2 - i \hat{\omega} \hat{\mathcal{M}} {m^2 \over (\lambda-3)} & \hat{\omega} \hat{k} & \text{\tiny $\nu = t$} \\
			& \hat{\omega} \hat{k} & \hat{\omega} \left( \hat{\omega} + i \hat{\mathcal{M}} {m^2 \over \lambda-3} \right) & \text{\tiny $\nu = x$} \\
		\end{block}
	\end{blockarray} ,
\end{aligned} \end{equation}
where we have introduced $\mathcal{C} \vcentcolon = {\lambda-3 \over \hat{\mathcal{M}}} \left[ {m^2 \over (\lambda-3)^2} \hat{\mathcal{M}}^2 + 1 \right]$. Note that we have simplified the order of subleading terms by assuming that $\hat{\omega} \sim \varepsilon$ and $\hat{\mathcal{M}} \sim \varepsilon^{-1}$. \\ \hspace*{1.75ex}
The expression above has poles given by \cref{7.8}, with $\tau_{\text{gap}}^{-1} \to r_h \mathcal{C}$ and $k^2 \to k^2 + r_h^2 m^2$. Hence, we expect the crossover associated with a ``pole collision structure" but with a second quasihydrodynamic relaxation mode in the place of hydrodynamic diffusion. However, unlike the previous section where $\tau_{\text{gap}}$ depended linearly on the coupling constant $\hat{\mathcal{M}}^{(\prime)}$, here $\tau_{\text{gap}} = 1/(r_h \mathcal{C})$ as a function of $\hat{\mathcal{M}}$ has a maximum at order $O (\varepsilon^{-1})$. This means that there are no relaxed sound modes in the scaling limit where $\hat{k} \sim \varepsilon \sim m$ since they arise for $\hat{k}^2 + m^2 \gg \mathcal{C}^2$. In particular, because $\hat{k}$ is fixed in this way, we cannot access the pole collision through the current analysis.

\paragraph{}Upon performing the aforementioned substitutions in the solutions \eqref{7.9} of \cref{7.8}, we obtain a pair of relaxation modes for $\hat{k}^2 + m^2 \ll \mathcal{C}^2$. When $1 < |\hat{\mathcal{M}}| <  \varepsilon^{-2}$ and $\hat{\mathcal{M}} \nsim \varepsilon^{-1}$, the dispersion relations are approximately given by
\begin{equation} \begin{aligned} \label{9.47}
	\hat{\omega} \approx - i {\hat{k}^2 + m^2 \over \mathcal{C}} \qquad \quad \text{and} \qquad \quad \hat{\omega} \approx - i \mathcal{C} + i {\hat{k}^2 + m^2 \over \mathcal{C}} \, .
\end{aligned} \end{equation} 

\paragraph{}The mode corresponding to the right-hand side relation acquires a finite gap, and is therefore absent from the low-energy spectrum, when either $|\hat{\mathcal{M}}| \lesssim 1$ or $|\hat{\mathcal{M}}| \gtrsim \varepsilon^{-2}$. In these cases, the left-hand side correlator in \eqref{9.45} is given respectively by
\begin{equation} \begin{aligned} \label{6.64}
	{\delta K_\mp^\mu \over \delta \mathcal{K}^\pm_\nu} & = {i r_h^{\lambda-3} / (\lambda-3) + O (\varepsilon^2) \over \hat{\omega} - \hat{\omega}_* \left( 1 + \hat{\mathcal{M}} \right) + O (\varepsilon^4)} 
	\begin{blockarray}{cccc}
		& \text{\tiny $\mu = t$} & \text{\tiny $\mu = x$} & \\
		\begin{block}{c(cc)c}
			& \hat{k}^2 + m^2 & \hat{\omega} \hat{k} & \text{\tiny $\nu = t$} \\
			& \hat{\omega} \hat{k} & \hat{\omega}^2 + i {m^2 \over \lambda-3} \hat{\omega}_* + i \hat{\mathcal{M}} {m^2 \over \lambda-3} (\hat{\omega} + \hat{\omega}_*) & \text{\tiny $\nu = x$} \\
		\end{block}
	\end{blockarray} ,
\end{aligned} \end{equation}
where we introduced $\hat{\omega}_* \vcentcolon = - i {\hat{k}^2 + m^2 \over \lambda-3}$, and
\begin{equation} \begin{aligned} \label{6.68b} 
	{\delta K_\mp^\mu \over \delta \mathcal{K}^\pm_\nu} & = {r_h^{\lambda-3} / \hat{\mathcal{M}} + O (\varepsilon^4) \over \hat{\omega} + \hat{\omega}_* \left( 1 - {(\lambda-3)^2 \over m^2 \hat{\mathcal{M}}} \right) + O (\varepsilon^4)} 
	\begin{blockarray}{cccc}
		& \text{\tiny $\mu = t$} & \text{\tiny $\mu = x$} & \\
		\begin{block}{c(cc)c}
			& \hat{\omega} + \hat{\omega}_* \left( 1 - {(\lambda-3)^2 \over m^2 \hat{\mathcal{M}}} \right) & - i {3-\lambda \over m^2} {\hat{\omega} \hat{k} \over \hat{\mathcal{M}}} & \text{\tiny $\nu = t$} \\
			&  - i {3-\lambda \over m^2} {\hat{\omega} \hat{k} \over \hat{\mathcal{M}}} & - \hat{\omega} - \hat{\omega}_* & \text{\tiny $\nu = x$} \\
		\end{block}
	\end{blockarray} .
\end{aligned} \end{equation}
Both these expressions followed from \cref{7.46}. Once again, we have simplified our presentation of the order of subleading terms by assuming that $\hat{\omega} \sim \varepsilon^2$ together with $\hat{\mathcal{M}} \sim O (1)$ and $\hat{\mathcal{M}} \sim O (\varepsilon^{-2})$ for the top and bottom equations, respectively. Note that both relaxation poles above agree (to leading order) with the left-hand side dispersion relation in \eqref{9.47}, when their regime of validity overlaps with $1 < |\hat{\mathcal{M}}| <  \varepsilon^{-2}$.

\paragraph{}Massive Hodge duality (horizontal arrows) and strong/weak-coupling duality (vertical arrows) act on the correlators \eqref{9.45} according to the following diagram
\begin{center}
	\begin{tikzpicture}
		\node (E) at (0,0) {$\braket{\mathcal{O}_a^{\mu A_2 ... A_p} \mathcal{O}_a^{\nu A_2 ... A_p}}_{\mathrm{R}}$};
		\node
		(F) at (5,0) {$\braket{\mathcal{O}_j^{\mu' A'_2 ... A'_{p'}} \mathcal{O}_j^{\nu' A'_2 ... A'_{p'}}}_{\mathrm{R}}$};
		\node
		(N) at (5,-1.8) {$\braket{\mathcal{O}_a^{\mu' A'_2 ... A'_{p'}} \mathcal{O}_a^{\nu' A'_2 ... A'_{p'}}}_{\mathrm{R}}$};
		\node[below=of E] (M) {$\braket{\mathcal{O}_j^{\mu A_2 ... A_p} \mathcal{O}_j^{\nu A_2 ... A_p}}_{\mathrm{R}}$};
		\draw[<->] (E)--(F) node [midway,above] {};
		\draw[<->] (F)--(N) node [midway,right] {};
		\draw[<->] (M)--(N) node [midway,below] {};
		\draw[<->] (E)--(M) node [midway,left] {};
	\end{tikzpicture} 
\end{center}
where $\mu , \nu , \mu' , \nu' \in \{ t , x \}$ and $p' = d - p$. The composition of both dualities implies that the pole of \eqref{6.71} has to be invariant under \eqref{9.44} followed by $\hat{\mathcal{M}}^\ast \to 1 / \hat{\mathcal{M}}$. The coefficient $\mathcal{C}$, in particular, is {a priori} constrained such that $\lambda$, $m^2$ and $\hat{\mathcal{M}}$ enter only through a specific combination:
\begin{equation} \begin{aligned}
	\mathcal{C} \left( \lambda , m^2 , \hat{\mathcal{M}} \right) = \mathcal{C} \left( {m^2 \hat{\mathcal{M}} \over \lambda-3} + {\lambda-3 \over \hat{\mathcal{M}}} \right) .
\end{aligned} \end{equation}

\subsubsection{Massless limit} \label{massless_limit}

\paragraph{}We end this section by noting how the massless correlators from \cref{masslesscorrelators} can be obtained from massive correlators in the massless limit, by which we mean taking $m^2 \to 0$ while holding fixed $\mathcal{M}$ and ${\mathcal{M}^\ast \over m^2}$ (the latter of which we denote by $\mathsf{M}$ to avoid clutter). \\ \hspace*{1.75ex}
Retarded correlators in massive and massless electric theories are then related according to
\begingroup
\renewcommand{\arraystretch}{1.4}
\begin{equation} \label{limitZa}
	\begin{array}{c}
		\qquad Z^{[\lambda , \mathcal{M}]} \dashrightarrow \bar{Z}^{[\lambda - 2 , {\mathcal{M} \over \lambda-3}]} \\ \hline
		\varDelta_\mp \braket{\mathcal{O}_j^{A_1 ... A_p} \mathcal{O}_j^{A_1 ... A_p}}_{\mathrm{R}} \xrightarrow[m^2 \to 0]{} - \braket{\bar{\mathcal{O}}_j^{A_1 ... A_p} \bar{\mathcal{O}}_j^{A_1 ... A_p}}_{\mathrm{R}} \\
		\varDelta_\mp \braket{\mathcal{O}_j^{\mu A_2 ... A_p} \mathcal{O}_j^{\nu A_2 ... A_p}}_{\mathrm{R}} \xrightarrow[m^2 \to 0]{} - \braket{\bar{\mathcal{O}}_j^{\mu A_2 ... A_p} \bar{\mathcal{O}}_j^{\nu A_2 ... A_p}}_{\mathrm{R}} \\
		\varDelta_\mp \braket{\mathcal{O}_j^{t x A_3 ... A_p} \mathcal{O}_j^{t x A_3 ... A_p}}_{\mathrm{R}} \xrightarrow[m^2 \to 0]{} 0 \, , \qquad \qquad \qquad \qquad \quad \;
	\end{array}
\end{equation}
\endgroup
where $\mu , \nu \in \{ t , x \}$, and for the magnetic theories we have 
\begingroup
\renewcommand{\arraystretch}{1.4}
\begin{equation} \label{limitZj}
	\begin{array}{c}
		\qquad \quad \; \; Z^{[\lambda , m^2 \mathsf{M}]} \dashrightarrow \bar{Z}^{[\lambda , (\lambda-3) \mathsf{M}]} \\ \hline
		\varDelta_\pm \braket{\mathcal{O}_a^{\mu A_2 ... A_p} \mathcal{O}_a^{\nu A_2 ... A_p}}_{\mathrm{R}} \xrightarrow[m^2 \to 0]{} \braket{\bar{\mathcal{O}}^a_{A_1 ... A_p} \bar{\mathcal{O}}^a_{A_1 ... A_p}}_{\mathrm{R}} \\
		\varDelta_\pm \braket{\mathcal{O}_a^{t x A_3 ... A_p} \mathcal{O}_a^{t x A_3 ... A_p}}_{\mathrm{R}} \xrightarrow[m^2 \to 0]{} \braket{\bar{\mathcal{O}}^a_{\mu A_2 ... A_p} \bar{\mathcal{O}}^a_{\nu A_2 ... A_p}}_{\mathrm{R}} \\
		\varDelta_\pm \braket{\mathcal{O}_a^{A_1 ... A_p} \mathcal{O}_a^{A_1 ... A_p}}_{\mathrm{R}} \xrightarrow[m^2 \to 0]{} 0 \, . \qquad \qquad \qquad \; \;
	\end{array}
\end{equation}
\endgroup
One can see this explicitly from the expressions for massive correlators that were given above. The leading parts of these when $m^2$ is small are displayed in appendix \ref{masslesslimit}. 

\section{Summary and outlook} \label{discussion}

\paragraph{}In this paper, we explored holographic descriptions of the low-energy dynamics of higher-form symmetries, both exact and approximate. We computed the large-$N$ thermal spectra of collective excitations, across a broad theory space parametrised by the dimension of the charges, the double-trace coupling and the scale of weak symmetry breaking,\footnote{These are encoded respectively in the rank of the differential forms used in the description, the value of the $\mathcal{M}$ couplings and the mass of the bulk field.} for a system being probed at a low energy-to-temperature ratio in $d$ spacetime dimensions. 
As this computation was performed in the probe limit, it is expected to provide insight into the dynamics of near-equilibrium states at low charge density. We have verified explicitly and in detail that hydrodynamics and its quasihydrodynamic extension (as presented in \cite{Grozdanov:2018fic}) provide the relevant descriptions of these spectra. We have also confirmed the dualities studied in \cite{Pinheiro:2026ihi} through their action on 2-point correlators.

\paragraph{}It follows naturally from the discussion of massless limits in \cref{massless_limit} that the holographic models describing an exactly or approximately conserved electric current admit a unified treatment. More precisely, one may consider a Proca field governed by the action \eqref{action_p2} with Robin boundary conditions, now retaining the $m^2=0$ case rather than excluding it.\footnote{Note that, although we assume $\lambda \neq 3$ when $m^2 \neq 0$, our results include the massless ${\bar{\lambda}} = 1 \Leftrightarrow \lambda = 3$ case.} With this perspective in mind, we conclude by summarising our results in the context of electric quantisation. \\ \hspace*{1.75ex}
Our unified model describes a strongly coupled fluid with a $(p{-}1)$-form chemical potential when 
\[m^2 = 0 \qquad \text{and} \qquad |\mathcal{M}| \lesssim T^{3-\lambda} , \] 
where $\mathcal{M}$ is the double-trace deformation coupling. In this regime, the low-energy spectra feature hydrodynamic diffusive modes in the longitudinal channel. Working in Fourier space, where rotational symmetry was used to align the wavevector according to $\vec{k} = k \hat{x}$ (without loss of generality), the longitudinal correlators are $\{ \braket{\mathcal{O}_j^{\mu A_2 \ldots A_p} \mathcal{O}_j^{\nu A_2 \ldots A_p}} \big\vert \mu,\nu = t , x \}$. Assuming a small, nonzero mass, the meromorphic correlators become
\begin{subequations} \label{low} \begin{align} 
	\label{4} \braket{\mathcal{O}_j^{\mu A_2 \ldots A_p} \mathcal{O}_j^{\nu A_2 \ldots A_p}} & \sim \left[ \hat{\omega} + i {\hat{k}^2 + m^2 \over \lambda-3} \left( 1 + \hat{\mathcal{M}} \right) + \ldots \right]^{-1} \\
	\label{5} \braket{\mathcal{O}_j^{t x A_3 \ldots A_p} \mathcal{O}_j^{t x A_3 \ldots A_p}} & \sim m^2 \left[ \hat{\omega} + i {\hat{k}^2 \over \lambda-1} + i {m^2 \over \lambda-3} \left( 1 + \hat{\mathcal{M}} \right) + \ldots \right]^{-1} .
\end{align} \end{subequations}
The poles of these correlators correspond to quasihydrodynamic modes. From \eqref{4} at $m^2 = 0$, one sees that for currents of degree higher than $d/2$ the corresponding fluid at low charge density is only stable against linearised perturbations if $\hat{\mathcal{M}} <  - 1$. \\ \hspace*{1.75ex}
There is a similar quasihydrodynamic regime for $|\mathcal{M}| \gtrsim T^{3-\lambda} / m^2$:
\begin{subequations} \label{high} \begin{align} 
	\braket{\mathcal{O}_j^{\mu A_2 \ldots A_p} \mathcal{O}_j^{\nu A_2 \ldots A_p}} & \sim \left[ \hat{\omega} - i {\hat{k}^2 + m^2 \over \lambda-3} \left( 1 - {(\lambda-3)^2 \over m^2 \hat{\mathcal{M}}} \right) + \ldots \right]^{-1} \\
	\label{7} \braket{\mathcal{O}_j^{A_1 \ldots A_p} \mathcal{O}_j^{A_1 \ldots A_p}} & \sim \left[ \hat{\omega} + i {\hat{k}^2 \over 5-\lambda} + i {m^2 \over 3-\lambda} \left( 1 - {(\lambda - 3)^2 \over m^2 \hat{\mathcal{M}}} \right) + \ldots \right]^{-1} .
\end{align} \end{subequations}
These correlators become analytic (in the low-frequency, low-wavenumber region of the complex $\omega \,$--$\, k$ plane) when $\mathcal{M} \to \infty$. As a result, this regime does not turn hydrodynamic and is indeed absent for $m^2 = 0$. Note that the dispersion relations of the low-$\mathcal{M}$ and high-$\mathcal{M}$ modes, associated with \eqref{low} and \eqref{high}, are related by
\begin{equation} \begin{aligned} \label{9}
	\lambda \to 6 - \lambda \qquad \text{and} \qquad \hat{\mathcal{M}} \to - {(\lambda - 3)^2 \over m^2 \hat{\mathcal{M}}} ,
\end{aligned} \end{equation}
which is the composition of the strong/weak-coupling duality with the massive Hodge duality --- cf. \cref{9.44}. As both dualities interchange the quantisation scheme, their composition maps back to electric quantisation.

\paragraph{}Equations \eqref{5} and \eqref{7} fully account for the low-energy excitations associated with $\langle{\mathcal{O}_j^{t x A_3 \ldots A_p} \mathcal{O}_j^{t x A_3 \ldots A_p} \rangle}$ and $\langle{\mathcal{O}_j^{A_1 \ldots A_p} \mathcal{O}_j^{A_1 \ldots A_p} \rangle}$, since they remain valid in the intermediate regime where
\[T^{3-\lambda} < |\mathcal{M}| < T^{3-\lambda} / m^2 .\]  
However, this does not apply to $\langle{\mathcal{O}_j^{\mu A_2 \ldots A_p} \mathcal{O}_j^{\nu A_2 \ldots A_p} \rangle}$ and we see new quasihydrodynamic modes emerging at intermediate-$\mathcal{M}$. As illustrated in figure \ref{crossover}, this sector is characterised by a `relaxed diffusion-to-sound crossover' which follows from poles at
\begin{equation} \begin{aligned}
	\hat{\omega} \left( \hat{\omega} + i \mathcal{C} \right) = \hat{k}^2 + m^2 + \ldots \quad \text{where} \quad \mathcal{C} = {\lambda - 3 \over \hat{\mathcal{M}}} \left[ {m^2 \over (\lambda - 3)^2} \hat{\mathcal{M}}^2 + 1 \right] .
\end{aligned} \end{equation}
This equation is solved by a pair of dispersion relations describing modes that relax when $\mathcal{C}^2 \gg \hat{k}^2 + m^2$ and propagate when $\mathcal{C}^2 \ll \hat{k}^2 + m^2$.
\begin{figure}[t]
	\pgfmathsetmacro{\figH}{3.5}
	\pgfmathsetmacro{\figV}{1.3}
	\pgfmathsetmacro{\figHH}{\figH+1.0}
	\centering
	\scalebox{1}{
		\begin{tikzpicture}[>=stealth',shorten >=1pt,auto]
			\node[scale=1.7] at (0.3*\figH,-0.5*\figV) {\color{black} \tiny $\hat{\omega} = - i {\mathcal{C}^2 - m^2 \over \mathcal{C}} + i {\hat{k}^2 \over \mathcal{C}} + \dots$};
			\node[scale=1.7] at (0.3*\figH,1.5*\figV) {\color{black} \tiny $\hat{\omega} = - i {m^2 \over \mathcal{C}} - i {\hat{k}^2 \over \mathcal{C}} + \dots$};
			\node[scale=1.7] at (\figH+\figHH,0.45*\figV) {\color{black} \tiny $\hat{\omega} = \pm \hat{k} - i {\mathcal{C} \over 2} + \dots$};
			\node[scale=1.5,align=center] at (\figH+0.5,0.55*\figV) {\color{black} \tiny pole \\[-1.5ex] \color{black} \tiny collision};
			\node (x1) at (0,0) {};
			\node (x2) at (\figH,0) {};
			\node (y1) at (0,\figV) {};
			\node (y2) at (\figH,\figV) {};
			\node (x3) at (\figHH,0) {};
			\node (x4) at (\figH+\figHH,0) {};
			\node (y3) at (\figHH,\figV) {};
			\node (y4) at (\figH+\figHH,\figV) {};
			\node (z1) at (0,2*\figV) {};
			\node (z2) at (\figH+\figHH,2*\figV) {};
			\node (zz1) at (0,-\figV) {};
			\node (zz2) at (\figH+\figHH,-\figV) {};
			\node (Xaxis1) at (-0.1,-1.5*\figV) {};
			\node (Xaxis2) at (\figH+\figHH+0.2,-1.5*\figV) {};
			\node[scale=1.7] at (\figH+\figHH,-1.5*\figV-0.4) {\tiny $\hat{k}^2$};
			\node (Yaxis1) at (\figH+0.5,-\figV-1.05) {};
			\node (Yaxis2) at (\figH+0.5,-\figV-0.25) {};
			\node[scale=1.7] at (\figH+0.6,-\figV-1.15) {\tiny $\mathcal{C}^2 - m^2$};
			
			\path[draw=black,-]
			(x1) edge node {} (x2)
			(y1) edge node {} (y2);
			\path[draw=red,-]
			(z1) edge node {} (z2);
			\path[draw=blue,-]
			(zz1) edge node {} (zz2);
			
			\path[->,thick]
			(Xaxis1) edge node {} (Xaxis2);
			
			\path[-,thick]
			(Yaxis1) edge node {} (Yaxis2);
			
			\draw[draw=black,decorate, decoration={snake}] 
			(x3) -- (x4)
			(y3) -- (y4);
			
			\draw[draw=black,dashed,thick] (\figH+0.5,\figV/2) circle (0.9cm);
		\end{tikzpicture}
	}
	\caption{\label{crossover}schematic representation of the spectrum's dependence on wavenumber, in the intermediate-$\mathcal{M}$ regime: low-energy modes associated with $\mathcal{O}_j^{A_1 \ldots A_p}$ (red), $\mathcal{O}_j^{\mu A_2 \ldots A_p}$ (black) and $\mathcal{O}_j^{t x A_3 \ldots A_p}$ (blue), for fixed $A_1 ... A_p$. Straight and wiggly lines correspond to relaxing and propagating modes, respectively.}
\end{figure}
Importantly, $\mathcal{C}^2$ has a minimum at $\mathcal{M}^2 = {(\lambda - 3)^2 \over m^2}$, for which $\mathcal{C}^2 = 4 m^2$. This, in particular, implies that only relaxation modes are accessible when probing the defect scale. 

\paragraph{}We remarked in the introduction that, for the models under consideration, quasihydrodynamics is necessary ``even when the higher-form symmetry is exact, provided the deformations are strong". We are now in a position to offer an interpretation regarding the origin of quasihydrodynamic modes when $m^2 = 0$. First, note that in the massless case the high-$\mathcal{M}$ regime above is absent, with the intermediate-$\mathcal{M}$ regime taking its place instead. \\ \hspace*{1.75ex}
From the perspective of the undeformed theory with an electric symmetry, taking $\mathcal{M}$ sufficiently large appears to produce (at least partially) an emergent approximate magnetic symmetry dual to the original one, leading to the electromagnetic structure discussed in \cref{electromag}. Conversely, from the viewpoint of the maximally deformed ($\mathcal{M} \to \infty$) theory, which enjoys two exact higher-form symmetries, turning on a nonzero inverse coupling explicitly breaks the magnetic symmetry. As long as $\mathcal{M}^{-1}$ remains sufficiently small, this breaking is weak and the resulting low-energy excitations are relaxed sound modes. \\ \hspace*{1.75ex}
One might then expect that, as the inverse coupling increases and the breaking becomes stronger, these quasihydrodynamic modes eventually disappear from the low-energy spectrum. Instead, before this can occur, a hydrodynamic (diffusive) mode protected by the electric symmetry emerges from the pole collision. This mode persists in the low-energy spectrum irrespective of how large $\mathcal{M}^{-1}$ becomes.

\paragraph{}Although all bulk computations in this work were performed in the probe limit, we expect our results to provide useful guidance for future, more phenomenologically oriented studies in which the full dynamics --- including fluctuations of the stress tensor, energy, and momentum --- are taken into account.\footnote{See \cref{intro} for examples of approximate higher-form symmetries found in nature.} One possible direction is the extension of the fluid-gravity correspondence \cite{Bhattacharyya:2007vjd,Rangamani:2009xk,Hubeny:2011hd} to incorporate such symmetry-breaking patterns. In this context, it would be particularly interesting to generalise the fluid-gravity description of viscoelastic crystals \cite{Davison:2025sze}, itself based on the higher-form model of \cite{Grozdanov:2018ewh}, so as to include the dynamics of melting through dislocation formation \cite{JMKosterlitz_1972,JMKosterlitz_1973,PhysRevLett.41.121,PhysRevB.19.2457,PhysRevB.19.1855} (see \cite{Beekman:2016szb,Beekman:2017brx} for a more modern account in the context of quantum liquid crystals).

\acknowledgments
I am very grateful to Arpit Das and Richard A. Davison for many helpful comments. This work was supported by an EPSRC Doctoral Training Partnership Award.

\appendix

\section{Exterior-calculus conventions} \label{conventions_append}

\paragraph{}We lay out the conventions adopted for exterior calculus. While $\Omega^p (M)$ denotes the space of smooth differential $p$-forms on a manifold $M$, $\Omega^p$ without an explicit manifold refers to $p$-forms on the $d$-dimensional physical spacetime.
The ``{components}" of $\omega = \omega_{\mu_1 ... \mu_p} \mathrm{d} x^{\mu_1} \wedge \ldots \wedge \mathrm{d} x^{\mu_p} \in \Omega^p$, are $ \omega_{\mu_1 ... \mu_p}$. The Hodge Star $\ast$ map associated with the metric $\gamma$ is such that the components of $\ast \omega$ are given by
\begin{equation} \begin{aligned} 
	(\ast \omega)_{\mu_1 ... \mu_{d-p}} & = {\epsilon_{\mu_1 ... \mu_{d-p} \nu_1 ... \nu_p} \omega^{\nu_1 ... \nu_p} \over p!} \, .
\end{aligned} \end{equation} 
where $\epsilon$ is the volume form ($\epsilon_{1 ... d} = \sqrt{|\gamma|}$).
Normalisation of the exterior derivative is such that $(\mathrm{d} \omega)_{\mu_0 ... \mu_p} = \partial_{[ \mu_0} \omega_{\mu_1 ... \mu_p]}$. \\ \hspace*{1.75ex}
Moving to the holographic bulk, we have the Hodge Star $\star$ associated with $g$ such that 
\begin{equation} \begin{aligned} 
	(\star \omega)_{a_0 ... a_{d-p}} & = {\tilde{\epsilon}_{a_0 ... a_{d-p} b_1 ... b_p} \omega^{b_1 ... b_p} \over p!} \, ,
\end{aligned} \end{equation} 
where $\omega \in \Omega^p (\mathbb{B})$ and $\tilde{\epsilon}$ is the volume form ($\tilde{\epsilon}_{r 1 ... d} = \sqrt{|g|}$).
We also define the adjoint exterior derivative $\mathrm{d}^\dagger$ according to
\begin{equation} \begin{aligned} 
	(\mathrm{d}^\dagger \omega)_{a_2 ... a_p} \vcentcolon = {(-1)^{p (d-p)} \over (d+1-p)!} ({\star \mathrm{d} \star} \omega)_{a_2 ... a_p} \, . 
\end{aligned} \end{equation} 

\section{Higher-form symmetries} \label{higherformsym}

\paragraph{}In this appendix, we provide some background on higher-form symmetries and motivate \cref{nonconseq}, according to which $\mathcal{O}$ ceases to be locally conserved where the defect current $\tilde{\mathcal{O}}$ is non-null. To this end, we begin by considering a $(p{-}1)$-form continuous symmetry, in which case there is a $p$-form Noether current that is co-closed,
\begin{equation} \begin{aligned} \label{conseq}
	\mathrm{d} \ast \mathcal{O} = 0 \, , 
\end{aligned} \end{equation}
and use this to develop the intuition for introducing defects into the system. Under this symmetry, the charge $Q (M)$, obtained by integrating $\ast \mathcal{O}$ over codimension-$p$ manifolds $M$, is a topological operator \cite{Bhardwaj:2023kri}. The charge density $\rho$ is the projection of $\mathcal{O}$ along the normal $n = g^{t \mu} \partial_\mu$ to codimension-1 spatial slices $\Sigma$. Hence, $\rho = \iota_n \mathcal{O}$ is a $(p{-}1)$-form and its integral hypersurfaces are the $(p{-}1)$-dimensional objects ``counted" at a time $t$ by the charge $Q (M)$, where $M \subset \Sigma_t$.\footnote{The interior product is defined such that $(\iota_X Y)_{\mu_2 \ldots \mu_q} = X^{\mu_1} Y_{\mu_1 ... \mu_q}$, where $X = X^\mu \partial_\mu$ and $Y \in \Omega^q$.} (The norm of $\rho$ encodes the density of integral hypersurfaces at a given point in space). To be specific, 
\begin{equation} \begin{aligned} 
	Q (M) \equiv \int_{M \subset \Sigma_t} \ast \mathcal{O} 
\end{aligned} \end{equation}
counts the intersections between $M$ and the integral hypersurfaces of $\rho$. The symmetry guarantees that the charge is conserved according to $Q (M')  = Q (M)$, where $M' \subset \Sigma_{t'}$, which can be shown by integrating \cref{conseq} over a $(d-p+1)$-dimensional manifold bounded (at least in part) by $M$ and $M'$.

\paragraph{}The $p$-dimensional worldvolumes along which the charged objects propagate have no $(p{-}1)$-dimensional junctions or boundaries. This follows from the relativistic nature of our conservation equation, that when integrated over any manifold yields only boundary contributions. In particular, the integral hypersurfaces of $\rho$ cannot terminate or meet at $(p{-}2)$-dimensional junctions --- being either closed or infinitely extended manifolds. This can be seen explicitly if we consider the projection along the normal $n$ of the conservation equation \eqref{conseq}:
\begin{equation} \begin{aligned} \label{spaceconseq}
	\iota_n (\mathrm{d} \ast \mathcal{O}) = - \mathrm{d} \ast \rho = 0 \, . 
\end{aligned} \end{equation}
The second equality is nothing less than a continuity equation for the charge density living on $\Sigma$. In realistic physical systems, however, higher-form symmetries are seldom realised exactly \cite{Armas:2023tyx}. In particular, if charged objects contain lower-dimensional imperfections then \cref{spaceconseq} has necessarily to be modified:
\begin{equation} \begin{aligned} \label{0.7}
	{\mathrm{d} \ast \rho \over d-p+1} = \ell \ast \tilde{\rho} \, .
\end{aligned} \end{equation}
Given this equation, where we have introduced the density $\tilde{\rho}$ associated with the $(p{-}1)$-form $\tilde{\mathcal{O}}$, the integral hypersurfaces of $\rho$ can have junctions or boundaries at the integral hypersurfaces of $\tilde{\rho} = \iota_n \tilde{\mathcal{O}}$, referred to as \textit{imperfections}. \Cref{0.7} is the projection along $n$ of the covariant (`non-conservation') equation
\begin{equation} \begin{aligned} \label{nonconseq_append}
	\mathrm{d} \ast \mathcal{O} = \ell \ast \tilde{\mathcal{O}} \, ,
\end{aligned} \end{equation}
which replaces \cref{conseq}. Conservation of the {defect current} $\tilde{\mathcal{O}}$ follows from the adjoint exterior derivative of \eqref{nonconseq_append}, implying the imperfections are charged under a $(p{-}2)$-form continuous symmetry and sweep out smooth, boundary-free worldvolumes --- the \textit{defects}. The $p$-dimensional worldvolumes swept by the original charged objects, on the other hand, are no longer boundary-free manifolds. In particular, timelike boundaries coincide with $(p{-}1)$-dimensional defects. What about spacelike boundaries then? These are associated with charge creation/destruction at a given moment. Although our initial motivation came from defects, the covariant formulation reveals two distinct sources of symmetry breaking. 
	
\section{Hydrodynamic solutions to equations of motion} \label{appendD}

\paragraph{}The primary purpose of this appendix is to provide computational details relevant to \cref{hydrolim}. Specifically, after solving the ${\bar{\lambda}}_{\text{eff}} = 1$ and $\lambda_{\text{eff}} = 1,3,5$ systems in a gradient expansion, we substitute these solutions into the corresponding ingoing wave conditions. Before doing so, we present additional details on how ingoing boundary conditions are imposed for solutions of the $\lambda_{\text{eff}} = 3$ system of equations.

\paragraph{}The near horizon behaviour dictated by the $\lambda_{\text{eff}} = 3$ system is given (unlike the systems corresponding to ${\bar{\lambda}}_{\text{eff}} = 1 , 3$ and $\lambda_{\text{eff}} = 1, 5$) by a set of coupled ODEs. To start, one can use \cref{24b} to eliminate the field $\mathcal{H}$ in \cref{24a}. After some manipulation, one sees that the dynamics of $\sqrt{|g|} \mathcal{F}^r$ and $\mathcal{F}_{x}$ is determined by
\begin{subequations} \begin{align}
	r^\lambda \partial_{r} \left( r^{4-\lambda} f (r) \partial_r \left( \sqrt{|g|} \mathcal{F}^r \right) \right) - f (r)^{-1} \left( \partial_{t}^2 + m^2 r^2 f (r) - f (r) \partial_{x}^2 \right) \sqrt{|g|} \mathcal{F}^r & = - r^\lambda f'(r) \partial_x \mathcal{F}_x \\
	r^{3} \partial_{r} \left( f (r) r^{\lambda-2} \partial_{r} \mathcal{F}_{x} \right) - f (r)^{-1} r^{\lambda-3} \left( \partial_{t}^2 - m^2 f (r) r^2 - f (r) \partial_x^2 \right) \mathcal{F}_{x} & = - 2 \partial_{x} \sqrt{|g|} \mathcal{F}^r .
\end{align} \end{subequations}
These equations admit solutions with ingoing behaviour at the horizon:
\begin{equation} \begin{aligned} \label{B.2}
	\{ \sqrt{|g|} \mathcal{F}^r , \mathcal{F}_{x} \} & \propto [ f'(r_h) (r-r_h)]^{- i \omega \over 4 \pi T} \left( 1 + O(r-r_h) \right)  ;
\end{aligned} \end{equation}
such that \cref{6.13} applies to $Y \in \{ \sqrt{|g|} \mathcal{F}^{r} , \mathcal{F}_{x} \}$.

\paragraph{}We proceed to solve the ${\bar{\lambda}}_{\text{eff}} = 1$ and $\lambda_{\text{eff}} = 1 , 3 , 5$ systems. The following formula will be useful:
\begin{equation} \begin{aligned} \label{formula}
	\int \mathrm{d} r {h (r) \over f (r)} & = \int \mathrm{d} r {h (r) \over f (r)} \left( 1 - {h (r_h) f' (r) \over h (r) f' (r_h) } \right) + {r_h^2 h (r_h) \over 4 \pi T} \ln f (r) \, ;
\end{aligned} \end{equation}
where $h (r)$ is some function that is analytic at the horizon and we used that $4 \pi T = r_h^2 f'(r_h)$.
\\[2ex] \noindent ${\boxed{{\bar{\lambda}}_{\text{eff}} = 1}}$ \\[1ex] \indent
Departing from the ingoing-wave condition \eqref{6.13} for $Y =\sqrt{|g|} F^{r t}$, we instead consider
\begin{equation} \begin{aligned} \label{6.14b} 
	\Xi (F_{t x}) & = {\partial_t \Gamma (F_{t x}) \over 4 \pi T} \, ,
\end{aligned} \end{equation}
which is equivalent by virtue of equations \eqref{eq:2.3}. We then seek to express this equation in terms of boundary fields by substituting $F_{t x}$ on-shell. This requires that we solve the ${\bar{\lambda}}_{\text{eff}} = 1$ system for $F_{t x}$. Hence, we start by integrating the non-radial components of \cref{eq:2.3a} and \cref{eq:2.3b}:
\begin{subequations} \begin{align} 
	F_{t x} & = \beta_{t x} + \partial_{t} \int \mathrm{d} r {\sqrt{|g|} F^{r x} \over f (r) r^{\bar{\lambda}}} + \partial_{x} \int \mathrm{d} r {\sqrt{|g|} F^{r t} \over r^{\bar{\lambda}}} \\
	\sqrt{|g|} F^{r t} & = J^{t} - \partial_{x} \int \mathrm{d} r {r^{{\bar{\lambda}}-4} \over f (r)} F_{t x} \\
	\sqrt{|g|} F^{r x} & = J^{x} + \partial_{t} \int \mathrm{d} r {r^{{\bar{\lambda}}-4} \over f (r)} F_{t x} \, .
\end{align} \end{subequations}
Recall that our convention for indefinite integrals is such that a function $g (r) \vcentcolon = \int \mathrm{d} r g' (r)$ has no $r$-independent term when expanded as $r \to \infty$. Substituting the bottom equations in the top one, we obtain
\begin{equation} \begin{aligned} \label{6.16}
	F_{t x} & = \beta_{t x} + \partial_{t} J^{x} \int \mathrm{d} r {r^{-{\bar{\lambda}}} \over f (r)} + \partial_{x} J^{t} {r^{1-{\bar{\lambda}}} \over 1-{\bar{\lambda}}} + O (\omega^2 , k^2) F_{t x} \, .
\end{aligned} \end{equation}
Using \cref{6.14b}, we have
\begin{equation} \begin{aligned} 
	\Gamma (F_{t x}) & = \beta_{t x} + \partial_{x} J^{t} {r_h^{1-{\bar{\lambda}}} \over 1-{\bar{\lambda}}} + O (\omega , k^2) F_{t x} \\
	& = {i 4 \pi T \over \omega} \Xi (F_{t x}) = J^{x} r_h^{2-{\bar{\lambda}}} + O (\omega) F_{t x} \, , 
\end{aligned} \end{equation}
where the top and bottom line originate from the analytic (near-horizon) term and the logarithmic divergence in \cref{6.16}, respectively. We also used \eqref{formula} for $h (r) = r^{-{\bar{\lambda}}}$. 
\\[2ex]\noindent ${\boxed{\lambda_{\text{eff}} = 5,1}}$ \\[1ex] \indent
Using \cref{23,26}, one can see that the ingoing wave conditions \eqref{6.13} for $Y = \mathcal{F}_{A_1 ... A_p}$ and $Y = \sqrt{|g|} \mathcal{H}^{r t x}$ are equivalent respectively to
\begin{subequations} \begin{align} 
	\label{210a} \Xi (\sqrt{|g|} \mathcal{H}^{r}) & = {\partial_t \Gamma (\sqrt{|g|} \mathcal{H}^{r}) \over 4 \pi T} \\
	\label{223a} \Xi (\mathcal{F}_{t x}) & = {\partial_t \Gamma (\mathcal{F}_{t x}) \over 4 \pi T} \, .
\end{align} \end{subequations}
In order to rewrite \cref{210a,223a} using boundary fields, we must first solve the massive $\lambda_{\text{eff}} = 5,1$ systems of equations for $\sqrt{|g|} \mathcal{H}^{r}$ and $\mathcal{F}_{t x}$. We start by integrating \cref{23a} and the radial component of \cref{23b} of the $\lambda_{\text{eff}} = 5$ system, yielding
\begin{subequations} \begin{align}
	\sqrt{|g|} \mathcal{H}^{r} & = (3-\lambda) K^{\mp} + \partial_{t}^2 \int \mathrm{d} r {r^{\lambda-6} \over f (r)} \mathcal{F}_{A_1 ... A_p} - \partial_{x}^2 \int \mathrm{d} r r^{\lambda-6} \mathcal{F}_{A_1 ... A_p} + m^2 \int \mathrm{d} r r^{\lambda-4} \mathcal{F}_{A_1 ... A_p} \\
	\mathcal{F}_{A_1 ... A_p} & = K^{\pm} + \int \mathrm{d} r {r^{2-\lambda} \over f (r)} \sqrt{|g|} \mathcal{H}^{r} \, .
\end{align} \end{subequations}
Note that integration constants have been identified with boundary fields by comparison with the solutions from \cref{p2setup} and we have used the non-radial components of \cref{23b} to write $\mathcal{H}_{\mu}$ in terms of $\mathcal{F}_{A_1 ... A_p}$. Substituting the bottom equation in the top one results in
\begin{equation} \begin{aligned} \label{6.23}
	\sqrt{|g|} \mathcal{H}^{r} & = (3-\lambda) K^{\mp} + \partial_{t}^2 K^{\pm} \int {r^{\lambda-6} \over f (r)} - \partial_{x}^2 K^{\pm} {r^{\lambda-5} \over \lambda-5} + m^2 K^{\pm} {r^{\lambda-3} \over \lambda-3} + O (m^2 , \omega^2 , k^2) \mathcal{H}^{r} \, .
\end{aligned} \end{equation}
Hence we can write, using \cref{210a},
\begin{equation} \begin{aligned} 
	\Gamma (\sqrt{|g|} \mathcal{H}^{r}) & = (3-\lambda) K^{\mp} - \partial_{x}^2 K^{\pm} {r_h^{\lambda-5} \over \lambda-5} + m^2 K^{\pm} {r_h^{\lambda-3} \over \lambda-3} + O ( \omega , k^2 , m^2) \mathcal{H}^{r} \\ 
	& = {i 4 \pi T \over \omega} \Xi (\sqrt{|g|} \mathcal{H}^{r}) = \partial_{t} K^{\pm} r_h^{\lambda-4} + O (\omega) \mathcal{H}^{r} ,
\end{aligned} \end{equation}
where the top and bottom line come respectively from the analytic (near-horizon) term and the logarithmic divergence in \cref{6.23}. Note that we also used \eqref{formula} for $h (r) = r^{\lambda-6}$. \\ \hspace*{1.75ex}
We now turn to the $\lambda_{\text{eff}} = 1$ system and integrate the non-radial component of \cref{26a} and \cref{26b}:
\begin{subequations} \begin{align} 
	\mathcal{F}_{t x} & = K^{\pm}_{t x} - {\partial_{t}^2 \over m^2} \int \mathrm{d} r {r^{- \lambda} \over f (r)} \sqrt{|g|} \mathcal{H}^{r t x} + {\partial_{x}^2 \over m^2} \int \mathrm{d} r r^{-\lambda} \sqrt{|g|} \mathcal{H}^{r t x} - \int r^{2-\lambda} dr \sqrt{|g|} \mathcal{H}^{r t x} \\
	\sqrt{|g|} \mathcal{H}^{r t x} & = (\lambda-3) K^{\mp}_{t x} - m^2 \int \mathrm{d} r {r^{\lambda-4} \over f (r)} \mathcal{F}_{t x} \, ;
\end{align} \end{subequations}
where we used the radial components of \cref{26a} to write $\mathcal{F}^{r \mu}$ in terms of $\mathcal{H}^{r t x}$. Substituting in the top equation the bottom one, we find
\begin{equation} \begin{aligned} \label{6.30}
	{m^2 \over 3-\lambda} \mathcal{F}_{t x} = {m^2 \over 3-\lambda} K^{\pm}_{t x} + \partial_{t}^2 K^{\mp}_{t x} \int \mathrm{d} r {r^{- \lambda} \over f (r)} - \partial_{x}^2 K^{\mp}_{t x} {r^{1-\lambda} \over 1-\lambda} & + m^2 K^{\mp}_{t x} {r^{3-\lambda} \over 3-\lambda} \\ & + m^2 O (m^2 , \omega^2 , k^2) \mathcal{F}_{t x} \, .
\end{aligned} \end{equation}
Thus we can write, using \cref{223a},
\begin{equation} \begin{aligned} 
	{m^2 \over 3-\lambda} \Gamma (\mathcal{F}_{t x}) & = {m^2 \over 3-\lambda} K^{\pm}_{t x} - \partial_{x}^2 K^{\mp}_{t x} {r_h^{1-\lambda} \over 1-\lambda} + m^2 K^{\mp}_{t x} {r_h^{3-\lambda} \over 3-\lambda} + m^2 O (m^2 , \omega , k^2) \mathcal{F}_{t x} \\
	& = {i 4 \pi T m^2 \over \omega (3-\lambda)} \Xi (\mathcal{F}_{t x}) = \partial_{t} K^{\mp}_{t x} r_h^{2-\lambda} + m^2 O (\omega) \mathcal{F}_{t x} \, ,
\end{aligned} \end{equation}
where the top and bottom line come, respectively, from the analytic (near-horizon) term and the logarithmic divergence in \cref{6.30}. Note that we also used \eqref{formula} for $h (r) = r^{-\lambda}$.
\\[2ex]\noindent ${\boxed{\lambda_{\text{eff}} = 3}}$ \\[1ex] \indent
Above, we concluded from \eqref{B.2} that imposing ingoing boundary conditions at the horizon requires
\begin{subequations} \begin{align} 
	\label{232b} \Xi (\mathcal{F}_{x}) & = {\partial_t \Gamma (\mathcal{F}_{x}) \over 4 \pi T} \\
	\label{232a} \Xi (\sqrt{|g|} \mathcal{F}^{r}) & = {\partial_t \Gamma (\sqrt{|g|} \mathcal{F}^{r}) \over 4 \pi T} \, .
\end{align} \end{subequations}
These equations imply 
\begin{subequations} \begin{align} 
	\label{233a} \Xi (\mathcal{F}_{t}) & = {\partial_t \Gamma (\mathcal{F}_{t}) \over 4 \pi T} \\
	\label{233b} \Xi (\sqrt{|g|} \mathcal{H}^{r t}) & = {\partial_t \Gamma (\sqrt{|g|} \mathcal{H}^{r t}) \over 4 \pi T} \, ,
\end{align} \end{subequations}
where we used $(\mathrm{d}^\dagger \mathcal{F})^{A_2 ... A_p} = 0$, which is the adjoint derivative of \cref{24a}, to derive the top equation and the $r t A_2 ... A_p$-component of \cref{24b} for the bottom one. \\ \hspace*{1.75ex}
We are going to solve the $\lambda_{\text{eff}} = 3$ system for $\mathcal{F}_{t}$ and $\mathcal{F}_{x}$ and then rewrite the ingoing wave conditions \eqref{232b} and \eqref{233a} in terms of boundary fields. Although this still follows the same logic as for the other systems, it is a bit more involved and we therefore find convenient to keep \eqref{232a} and \eqref{233b} at hand.

\paragraph{}We start by integrating the radial components of \cref{24b}, such that
\begin{subequations} \begin{align} 
	\label{Ft} \mathcal{F}_{t} & = K^{\pm}_t + \partial_{t} \int \mathrm{d} r {\sqrt{|g|} \mathcal{F}^{r} \over r^\lambda f (r)} - \int \mathrm{d} r {\sqrt{|g|} \mathcal{H}^{r t} \over r^{\lambda-2}} \\
	\label{Fz} \mathcal{F}_{x} & = K^{\pm}_x + \partial_{x} \int \mathrm{d} r {\sqrt{|g|} \mathcal{F}^{r} \over r^\lambda f (r)} + \int \mathrm{d} r {\sqrt{|g|} \mathcal{H}^{r x} \over r^{\lambda-2} f (r)} \, .
\end{align} \end{subequations}
Given this, we must first solve for $\sqrt{|g|} \mathcal{H}^{r \mu}$ and $\sqrt{|g|} \mathcal{F}^{r}$. Hence, we integrate the non-radial components of \cref{24a} and $(\mathrm{d}^\dagger \mathcal{F})^{A_2 ... A_p} = 0$ (which follows from \eqref{24a}) thus obtaining
\begin{subequations} \begin{align} 
	\sqrt{|g|} \mathcal{H}^{r t} & = (\lambda-3) K^{\mp}_t - m^2 \int \mathrm{d} r {r^{\lambda-4} \over f (r)} \mathcal{F}_{t} - \partial_{x} \int \mathrm{d} r {r^{\lambda-6} \over f (r)} \partial_{[t} \mathcal{F}_{x]} \\
	\sqrt{|g|} \mathcal{H}^{r x} & = (3-\lambda) K^{\mp}_x + m^2 \int \mathrm{d} r r^{\lambda-4} \mathcal{F}_{x} + \partial_{t} \int \mathrm{d} r {r^{\lambda-6} \over f (r)} \partial_{[t} \mathcal{F}_{x]} \\
	\sqrt{|g|} \mathcal{F}^{r} & = X^{\mp} + \partial_{t} \int \mathrm{d} r {r^{\lambda-4} \over f (r)} \mathcal{F}_{t} - \partial_{x} \int \mathrm{d} r r^{\lambda-4} \mathcal{F}_{x} \, ,
\end{align} \end{subequations}
where the non-radial component of \cref{24b} was used to get rid of $\mathcal{H}_{t x}$. Let us rewrite the bottom two equations using \cref{Fz}:
\begin{subequations} \begin{align} 
	\sqrt{|g|} \mathcal{H}^{r x} & = (3-\lambda) K^{\mp}_x + m^2 K^{\pm}_x {r^{\lambda-3} \over \lambda-3} \notag \\
	& + O (m^2 k) X^{\mp} + O (m^2) K^{\mp}_x + O (\omega) \partial_{[t} \mathcal{F}_{x]} + m^2 O (k^2 , m^2) \mathcal{F}_{x} \\
	\label{7.39b} \sqrt{|g|} \mathcal{F}^{r} & = X^{\mp} - \partial_{x} K^{\pm}_x {r^{\lambda-3} \over \lambda-3} + O (k^2) X^{\mp} + O (k) K^{\mp}_x + O (\omega) \mathcal{F}_{t} + k O (k^2 , m^2 , \omega^2) \mathcal{F}_{x} \, .
\end{align} \end{subequations}
As it stands, \cref{Ft} can be written as
\begin{equation} \begin{aligned} \label{6.38}
	\mathcal{F}_{t} = & K^{\pm}_t + \partial_{t} X^{\mp} \int \mathrm{d} r {r^{-\lambda} \over f (r)} + r^{3-\lambda} K^{\mp}_t + O (\omega k^2) X^{\mp} + O (\omega^2 , m^2 , k^2) \mathcal{F}_{t} + O (\omega k) \mathcal{F}_{x} \, ,
\end{aligned} \end{equation}
while \cref{Fz} is given by
\begin{equation} \begin{aligned} \label{6.40}
	\mathcal{F}_{x} = K^{\pm}_x + \partial_{x} \left( X^{\mp} - \partial_{x} K^{\pm}_x {r^{\lambda-3} \over \lambda-3} \right) \int \mathrm{d} r {r^{-\lambda} \over f (r)} + \left( (3-\lambda) K^{\mp}_x + m^2 K^{\pm}_x {r^{\lambda-3} \over \lambda-3} \right) \int \mathrm{d} r {r^{2-\lambda} \over f (r)} \\
	+ k O (k^2 , m^2) X^{\mp} + O (k^2 , m^2) K^{\mp}_x + O (\omega k) \mathcal{F}_{t} + O (k^4 , \omega^2 , m^2 k^2 , m^4) \mathcal{F}_{x} \, .
\end{aligned} \end{equation}

\paragraph{}In \cref{Ft}, if one takes \cref{232a,233b} into account, then it follows that $f' (r_h) \Xi (\mathcal{F}_{t}) = \partial_t \Gamma (\sqrt{|g|} \mathcal{F}^{r}) r_h^{-\lambda}$ (where \eqref{formula} was used for $h (r) = r^{-\lambda}$). Hence, we can write using \cref{233a,7.39b}
\begin{equation} \begin{aligned} \label{7.42}
	{i 4 \pi T \over \omega} \Xi (\mathcal{F}_{t}) & = r_h^{2-\lambda} X^{\mp} - \partial_{x} K^{\pm}_x {r_h^{-1} \over \lambda-3} \\
	& + O (k^2) X^{\mp} + O (k) K^{\mp}_x + O (\omega) \mathcal{F}_{t} + k O (k^2 , m^2 , \omega^2) \mathcal{F}_{x} \, .
\end{aligned} \end{equation}
From \cref{6.38}, we have
\begin{equation} \begin{aligned} \label{7.43}
	\Gamma (\mathcal{F}_{t}) = K^{\pm}_t + r_h^{3-\lambda} K^{\mp}_t + O (\omega^2 , m^2 , k^2) \mathcal{F}_{t} + O (\omega k) \mathcal{F}_{x} + O (\omega) X^{\mp} \, .
\end{aligned} \end{equation}
From the logarithmic divergence in \cref{6.40}, we can write using \eqref{232b}
\begin{equation} \begin{aligned} \label{7.44}
	4 \pi T \Xi (\mathcal{F}_{x}) & = \partial_{x} X^{\mp} r_h^{2-\lambda} - \partial_{x}^2 K^{\pm}_x {r_h^{-1} \over \lambda-3} + (3-\lambda) K^{\mp}_x r_h^{4-\lambda} + K^{\pm}_x {r_h m^2 \over \lambda-3} \\
	& + k O (k^2 , m^2) X^{\mp} + O (k^2 , m^2) K^{\mp}_x + O (\omega k) \mathcal{F}_{t} + O (k^4 , \omega^2 , m^2 k^2 , m^4) \mathcal{F}_{x} \, ,
\end{aligned} \end{equation}
where we used \eqref{formula} for $h (r) = r^{-\lambda}$ and $h (r) = r^{2-\lambda}$. Lastly, from the analytic (near-horizon) term in \cref{6.40}, we have
\begin{equation} \begin{aligned} \label{7.45}
	- i \omega \Gamma (\mathcal{F}_{x}) = \partial_t K^{\pm}_x + O (\omega k) X^{\mp} + \omega O (k^2 , m^2) K^{\pm}_x + O (\omega) K^{\mp}_x + O (k \omega^2) \mathcal{F}_{t} \\
	+ \omega O (k^4 , \omega^2 , m^2 k^2 , m^4) \mathcal{F}_{x} \, .
\end{aligned} \end{equation}
Substituting \cref{7.42,7.43,7.44,7.45} in the ingoing wave conditions \eqref{232b} and \eqref{233a} results in the pair of equations \eqref{6.45}. Upon the use of the approximate conservation equation \eqref{6.57}, these have been written in terms of the dimensionless frequency and wavenumber $(\hat{\omega} , \hat{k})$ --- cf. \eqref{7.47}. This equation applies to the scaling limit where $\hat{\omega} \sim \varepsilon$, given that $\hat{k} \sim \varepsilon \sim m$. It is also useful to include the $\hat{\omega} \sim \varepsilon^2$ case:
\begin{equation} \begin{aligned} \label{7.46}
	r_h^{\lambda-3} \begin{pmatrix}
		i {m^2 \over 3-\lambda} + O (\varepsilon^4) & \hat{k} {m^2 \over (\lambda-3)^2} + O (\varepsilon^5) \\
		O (\varepsilon^5) & i {m^2 \over \lambda-3} \left( \hat{\omega} - i {\hat{k}^2 + m^2 \over \lambda-3} + O (\varepsilon^4) \right)
	\end{pmatrix}
	\begin{pmatrix}
		K^{\pm}_t \\
		K^{\pm}_x
	\end{pmatrix} & \\
	=
	\begin{pmatrix}
		\hat{\omega} + i {m^2 \over \lambda-3} + O (\varepsilon^4) & \hat{k} + O (\varepsilon^3) \\
		\hat{\omega} \hat{k} + O (\varepsilon^5) & \hat{k}^2 + m^2 + O (\varepsilon^4)
	\end{pmatrix} \begin{pmatrix}
		K^{\mp}_t \\
		K^{\mp}_x
	\end{pmatrix} & .
\end{aligned} \end{equation}

\section{Magnetic correlators of the self-dual sector}

\paragraph{}In this appendix, we collect the expressions analogous to \eqref{6.71}, \eqref{6.64} and \eqref{6.68b} for the magnetic quantisation of massive theories. By virtue of \cref{4.42b}, these allow us to derive the large-$N$ limit of $\braket{\mathcal{O}_a^{\mu A_2 ... A_p} \mathcal{O}_a^{\nu A_2 ... A_p}}_{\mathrm{R}}$.

\paragraph{}Let us start with the regime where $\varepsilon^2 < |\hat{\mathcal{M}}| <  1$ and $\hat{\mathcal{M}}^\ast \nsim \varepsilon$, for which the relevant expression is
\begin{equation} \begin{aligned} \label{6.70}
	{\delta K_\pm^\mu \over \delta \mathcal{K}^\mp_\nu} & = {r_h^{3-\lambda} / \hat{\mathcal{M}}^\ast + O (1) \over \hat{\omega} \left( \hat{\omega} + i \mathcal{C}^\ast \right) - \hat{k}^2 - m^2 + O (\varepsilon^3)} \\ &
	\begin{blockarray}{cccc}
		& \text{\tiny $\mu = t$} & \text{\tiny $\mu = x$} & \\
		\begin{block}{c(cc)c}
			& \hat{\omega} \left( \hat{\omega} + i \hat{\mathcal{M}}^\ast (\lambda-3) \right) & \hat{\omega} \hat{k}  & \text{\tiny $\nu = t$} \\
			& \hat{\omega} \hat{k}  & \hat{k}^2 + m^2 - i (\lambda-3) \hat{\omega} \hat{\mathcal{M}}^\ast & \text{\tiny $\nu = x$} \\
		\end{block}
	\end{blockarray} ,
\end{aligned} \end{equation}
where $\mathcal{C}^\ast \vcentcolon = {\lambda-3 \over \hat{\mathcal{M}}^\ast} \left[ {m^2 \over (\lambda-3)^2} + (\hat{\mathcal{M}}^\ast)^2 \right]$. We have assumed that $\hat{\omega} \sim \varepsilon$ and $\hat{\mathcal{M}}^\ast \sim \varepsilon$ in order to simplify subleading terms. When $|\hat{\mathcal{M}}^\ast| \lesssim \varepsilon^2$, we have instead 
\begin{equation} \begin{aligned} \label{6.63}
	{\delta K_\pm^\mu \over \delta \mathcal{K}^\mp_\nu} & = {i r_h^{3-\lambda} {3-\lambda \over m^2} + O (1) \over \hat{\omega} + \hat{\omega}_* \left( 1 - \hat{\mathcal{M}}^\ast {(\lambda-3)^2 \over m^2} \right) + O (\varepsilon^4)} \\ &
	\begin{blockarray}{cccc}
		& \text{\tiny $\mu = t$} & \text{\tiny $\mu = x$} & \\
		\begin{block}{c(cc)c}
			& \hat{\omega}^2 + i {m^2 \over \lambda-3} \hat{\omega}_* + i \hat{\mathcal{M}}^\ast (\hat{\omega} - \hat{\omega}_*) (\lambda-3) &\hat{\omega} \hat{k} & \text{\tiny $\nu = t$} \\
			& \hat{\omega} \hat{k} & \hat{k}^2 + m^2 & \text{\tiny $\nu = x$} \\
		\end{block}
	\end{blockarray} ,
\end{aligned} \end{equation}
where $\hat{\omega} \sim \varepsilon^2$ and $\hat{\mathcal{M}}^\ast \sim \varepsilon^2$. Lastly, when $1 \lesssim |\hat{\mathcal{M}}^\ast|$, we have
\begin{equation} \begin{aligned} \label{6.68a} 
	{\delta K_\pm^\mu \over \delta \mathcal{K}^\mp_\nu} & = {- r_h^{3-\lambda}  / \hat{\mathcal{M}}^\ast + O (\varepsilon^2) \over \hat{\omega} - \hat{\omega}_* \left( 1 + {1 \over \hat{\mathcal{M}}^\ast} \right) + O (\varepsilon^4)} 
	\begin{blockarray}{cccc}
		& \text{\tiny $\mu = t$} & \text{\tiny $\mu = x$} & \\
		\begin{block}{c(cc)c}
			& - \hat{\omega} + \hat{\omega}_* & {i \over \hat{\mathcal{M}}^\ast} {\hat{\omega} \hat{k}  \over \lambda-3} & \text{\tiny $\nu = t$} \\
			& {i \over \hat{\mathcal{M}}^\ast} {\hat{\omega} \hat{k}  \over \lambda-3} & \hat{\omega} - \hat{\omega}_* \left( 1 + {1 \over \hat{\mathcal{M}}^\ast} \right) & \text{\tiny $\nu = x$} \\
		\end{block}
	\end{blockarray} ,
\end{aligned} \end{equation}
where $\hat{\omega} \sim \varepsilon^2$ and $\hat{\mathcal{M}}^\ast \sim O (1)$.

\section{Massless limit} \label{masslesslimit}

\paragraph{}As stated at the close of \cref{massivecorrelators}, all massless correlators from \cref{masslesscorrelators} arise from the massless limit of massive correlators. Such a limit refers to sending $m^2$ to zero with $\mathcal{M}$ and ${\mathcal{M}^\ast \over m^2}$ held constant. Here we present the leading parts of \cref{6.60,6.61,6.63,6.64,6.70,6.71} when $m^2$ is small. Comparing these with the expressions in \cref{masslesscorrelators}, one can confirm the mappings \eqref{limitZj} and \eqref{limitZa} between correlators. 
\\[2ex] \noindent $\triangleright$ From \cref{6.60a,6.60b}:
\begin{subequations} \begin{align}  
	\label{6.66c} {m^2 \over 3-\lambda} {\delta K_{\pm} \over \delta \mathcal{K}^\mp} & = {m^2 \over 3-\lambda} {- i r_h^{4-\lambda} (\lambda - 3) + O ( \varepsilon^2) \over \omega + i {r_h^{-1} k^2 \over 5-\lambda} + O ( \varepsilon^2 \mathcal{M}^\ast , \varepsilon^4)} + O (m^4) \\
	\label{6.67c} (3-\lambda) {\delta K_\mp^{t x} \over \delta \mathcal{K}^\pm_{t x}} & = m^2 {- i r_h^{\lambda-2} + O (\varepsilon^2) \over \omega + i {r_h^{-1} k^2 \over \lambda-1} + O (\varepsilon^4 \mathcal{M} , \varepsilon^4)} + O (m^4) \, .
\end{align} \end{subequations}
$\triangleright$ From \cref{6.61a,6.61b}:
\begin{subequations} \label{6.66} \begin{align} 
	\label{6.66a} {m^2 \over 3-\lambda} {\delta K_{\pm}^{t x} \over \delta \mathcal{K}^\mp_{t x}} & = {- m^2 \over \mathcal{M}^\ast (\lambda-3)} {\omega + i {k^2 r_h^{-1} \over \lambda-1} + O (\varepsilon^4) \over \omega + i {k^2 r_h^{-1} \over \lambda-1} + i {m^2 \over \lambda-3} {r_h^{\lambda-2} \over \mathcal{M}^\ast} + O (\varepsilon^4 / \mathcal{M}^\ast , \varepsilon^4)} + O (m^2) \\
	\label{6.67a} (3-\lambda) {\delta K_{\mp} \over \delta \mathcal{K}^\pm} & = {\lambda-3 \over \mathcal{M}} {\omega + i {r_h^{-1} k^2 \over 5-\lambda} + O ( \varepsilon^4) \over \omega + i {r_h^{-1} k^2 \over 5-\lambda} + i {r_h^{4-\lambda} \over \mathcal{M} / (\lambda-3)} + O (\varepsilon^2 / \mathcal{M} , \varepsilon^4)} + O (m^2) \, .
\end{align} \end{subequations}
$\triangleright$ From \cref{6.63,6.64} (which are valid when $0 \lesssim \mathcal{M}^\ast \lesssim O (\varepsilon^2)$ and $0 \lesssim \mathcal{M} \lesssim O (1)$):
\begin{subequations} \begin{align} 
	\label{6.66b} {m^2 \over 3-\lambda} {\delta K_\pm^\mu \over \delta \mathcal{K}^\mp_\nu} & = {i r_h^{2-\lambda} + O (\varepsilon^2) \over \omega + i \left( {r_h^{-1} \over 3-\lambda} + {\mathcal{M}^\ast \over r_h^{\lambda-2}} {\lambda-3 \over m^2} \right) k^2 + O (\varepsilon^4)} 
	\begin{blockarray}{cccc}
		& \text{\tiny $\mu = t$} & \text{\tiny $\mu = x$} & \\
		\begin{block}{c(cc)c}
			& \omega^2 & \omega k & \text{\tiny $\nu = t$} \\
			& \omega k & k^2 & \text{\tiny $\nu = x$} \\
		\end{block}
	\end{blockarray} + O (m^2) \\
	\label{6.67b} (3-\lambda) {\delta K_\mp^\mu \over \delta \mathcal{K}^\pm_\nu} & = {- i r_h^{\lambda-4} + O (\varepsilon^2) \over \omega + i \left( {r_h^{-1} \over \lambda-3} + {\mathcal{M} / (\lambda-3) \over r_h^{4-\lambda}} \right) k^2 + O (\varepsilon^4)} 
	\begin{blockarray}{cccc}
		& \text{\tiny $\mu = t$} & \text{\tiny $\mu = x$} & \\
		\begin{block}{c(cc)c}
			& k^2 & \omega k & \text{\tiny $\nu = t$} \\
			& \omega k & \omega^2 & \text{\tiny $\nu = x$} \\
		\end{block}
	\end{blockarray} + O (m^2) \, .
\end{align} \end{subequations}
$\triangleright$ From \cref{6.70,6.71} (which are valid when $\mathcal{M}^\ast \approx O (\varepsilon)$ and $\mathcal{M} \approx O (\varepsilon^{-1})$):
\begin{subequations} \begin{align} 
	{m^2 \over 3-\lambda} {\delta K_\pm^\mu \over \delta \mathcal{K}^\mp_\nu} & = {- {m^2 \over \lambda-3} \big/ \mathcal{M}^\ast + O (\varepsilon^3) \over \omega \left( \omega + i {r_h^{\lambda-2} \over \mathcal{M}^\ast} {m^2 \over \lambda-3} \right) - k^2 + O (\varepsilon^3)} 
	\begin{blockarray}{cccc}
		& \text{\tiny $\mu = t$} & \text{\tiny $\mu = x$} & \\
		\begin{block}{c(cc)c}
			& \omega^2 & \omega k & \text{\tiny $\nu = t$} \\
			& \omega k & k^2 & \text{\tiny $\nu = x$} \\
		\end{block}
	\end{blockarray} + O (m^2) \\
	(3-\lambda) {\delta K_\mp^\mu \over \delta \mathcal{K}^\pm_\nu} & = {(\lambda-3) / \mathcal{M} + O (\varepsilon^3) \over \omega \left( \omega + i {r_h^{4-\lambda} \over \mathcal{M}} (\lambda-3) \right) - k^2 + O (\varepsilon^3)} 
	\begin{blockarray}{cccc}
		& \text{\tiny $\mu = t$} & \text{\tiny $\mu = x$} & \\
		\begin{block}{c(cc)c}
			& k^2 & \omega k & \text{\tiny $\nu = t$} \\
			& \omega k & \omega^2 & \text{\tiny $\nu = x$} \\
		\end{block}
	\end{blockarray} + O (m^2) \, .
\end{align} \end{subequations}

\bibliographystyle{JHEP}
\bibliography{references}

\end{document}